\documentclass[conference]{IEEEtran}
\IEEEoverridecommandlockouts
\usepackage{cite}
\usepackage{amsmath,amssymb,amsfonts}
\usepackage{algorithmic}
\usepackage{graphicx}
\usepackage{subcaption}
\usepackage[font=footnotesize,labelfont=bf]{caption}
\usepackage{textcomp}
\usepackage{algorithm2e}
\usepackage{xcolor}
\usepackage{hyperref}
\def\BibTeX{{\rm B\kern-.05em{\sc i\kern-.025em b}\kern-.08em
    T\kern-.1667em\lower.7ex\hbox{E}\kern-.125emX}}
\author{\IEEEauthorblockN{Tianxiang Lyu}
\IEEEauthorblockA{Graduate School of Medicine\\Juntendo University, Japan\\Email: t.lyu.sa@juntendo.ac.jp}
\and
\IEEEauthorblockN{Mitsuhisa Sato}
\IEEEauthorblockA{Faculty of Health Data Science\\Graduate School of Medicine\\Juntendo University\\RIKEN Center of Computational Science, Japan\\Email: m.sato.zy@juntendo.ac.jp}
\and
\IEEEauthorblockN{Shigeki Aoki}
\IEEEauthorblockA{Faculty of Health Data Science\\Graduate School of Medicine\\Juntendo University, Japan\\Email: saoki@juntendo.ac.jp}
\and
\IEEEauthorblockN{Ryutaro Himeno}
\IEEEauthorblockA{Faculty of Health Data Science\\Graduate School of Medicine\\Juntendo University, Japan\\Email: r.himeno.tx@juntendo.ac.jp}
\and
\IEEEauthorblockN{Zhe Sun}
\IEEEauthorblockA{Faculty of Health Data Science\\Graduate School of Medicine\\Juntendo University\\
RIKEN Center for Advanced Photonics, Japan\\Email: z.sun.kc@juntendo.ac.jp}}

\begin{document}

\title{CORTEX: Large-Scale Brain Simulator Utilizing Indegree Sub-Graph Decomposition on Fugaku Supercomputer\\
}

\maketitle

\begin{abstract}
We introduce CORTEX, an algorithmic framework designed for large-scale brain simulation. Leveraging the computational capacity of the Fugaku Supercomputer, CORTEX maximizes available problem size and processing performance. Our primary innovation, Indegree Sub-Graph Decomposition, along with a suite of parallel algorithms, facilitates efficient domain decomposition by segmenting the global graph structure into smaller, identically structured sub-graphs. This segmentation allows for parallel processing of synaptic interactions without inter-process dependencies, effectively eliminating data racing at the thread level without necessitating mutexes or atomic operations. Additionally, this strategy enhances the overlap of communication and computation. Benchmark tests conducted on spiking neural networks, characterized by biological parameters, have demonstrated significant enhancements in both problem size and simulation performance, surpassing the capabilities of the current leading open-source solution, the NEST Simulator. Our work offers a powerful new tool for the field of neuromorphic computing and understanding brain function.
\end{abstract}


\section{BACKGROUND AND MOTIVATION}
Recent years have seen an unprecedented accumulation of data at various levels of brain organisation, thanks in large part to extensive biological research exemplified by works such as \cite{amunts2020julich}, with a detailed mapping of the brain's cytoarchitecture. There will be a sure demand for integrating all of them into a unified model, providing a unified view on how brain works as a whole\cite{Churchland_1986}. However, the fundamental barriers are scale and complexity\cite{DeFelipe_2015}, making it hard to achieve a general comprehension by experimental or theoretical methods\cite{Markram_2015, Markram_2019}, dealing with 86 billion neurons in the human brain and the interactions between molecules, neurons, microcircuits, and brain areas. To transcend these barriers, brain simulation was developed to provide a new way to gain an overview of all the forests of datasets, with reorganizing and integrating them in the context of the whole brain. Optimistically, it seems possible to not only fill the gaps that experiments and theoretical analysis will never manage to fill, but also find ways through the forests by considering the ecosystem of the brain architecture\cite{Markram_2019}.\\
Enabled by the computational power of leading-edge supercomputers, which has been steadily and exponentially increasing for the past 20 years\cite{top_500}, brain simulation towards human scale is one of the most ambitious scientific challenges in the 21st century\cite{kandel_2013}, playing an indispensable role in the investigation of the multi-scale brain\cite{Markram_2019}. The goal of brain simulation is to achieve a dense digital reconstruction of brain dynamics from experimental datasets and the most fundamental principles we can isolate to understand, linking the multiple layers from cells, circuits, areas to brain function and behavior, in order to make progress systematic and understanding tractable\cite{Markram_2019}.\\
However, due to the sparsity of brain architecture in both spatial and temporal domains, many existing algorithms and libraries\cite{metis_1998} can't be directly applied in brain simulation for optimizations. What's worse, in some simulators, mutexes or atomic operations\cite{thu_snn_1, thu_snn_2} are introduced, in order to avoid data racing at the thread level, but resulting in poor performance. At last, the communication overhead might be unbearable in large-scale simulation, because the ratio of computation and communication in brain simulation is much lower than primary HPC applications.\\
Fortunately, most instances of interest are far from the worst-case scenario and the performance of simulation can have orders of magnitude improvements compared with the naive approach, by focusing on the unique properties of brain architecture. Based on directed graph\cite{Deo_1974}, the main technical contribution of our paper is Indegree Sub-Graph Decomposition, along with a suite of parallel algorithms, which can not only perform domain decomposition of entire brain architecture among all processes with maximized problem size, but also avoid data racing at the thread level without any mutex or atomic operation. Having achieved impressive improvements in various aspects, CORTEX represents as a powerful brain simulation framework on leading-edge supercomputers\cite{fugaku}.\\
\subsection{Brain Architecture and Spiking Neural Networks}
\begin{figure}[h!]
	\centering
    \includegraphics[width=0.4\textwidth]{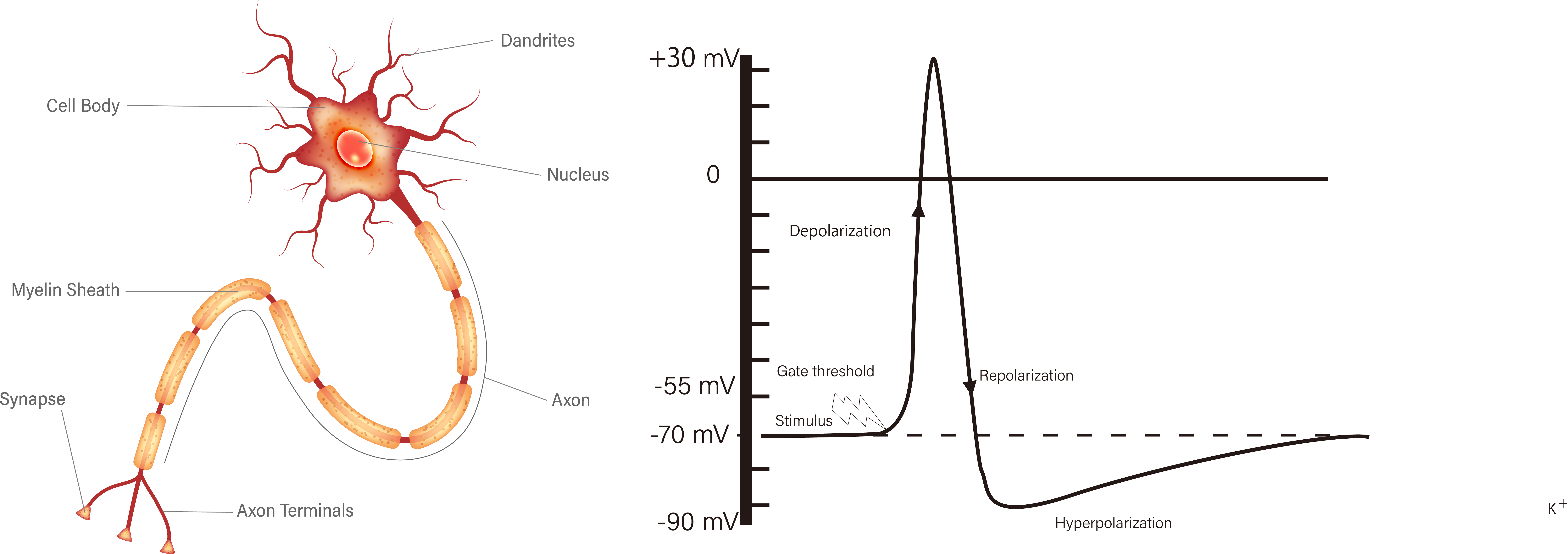}
    \caption{Schematic Representation of A Neuron and Its Spike: Illustrating the basic elements of (SNNs), where neurons are connected through synapses, forming a complex network. The "spike" serves as the primary unit of information processing in SNNs.}
    \label{fig:neuron_spike}
\end{figure}

In this section, a brief introduction about brain architecture consists of neurons and synapses fundamentally. More detailed reviews on neuron models have been published elsewhere\cite{neural_dynamics}. A neuron is a basic unit in the brain, connected with others through synapses as Fig.~\ref{fig:neuron_spike} shows. A "spike", also known as a neuronal spike or action potential, is the fundamental unit of information transfer in the nervous system. When a neuron is sufficiently stimulated, it generates a rapid, transient change in voltage. This phenomenon is referred to as a spike. All these neurons and synapses are forming a whole, which can be regarded as spiking neural networks (SNNs).\\
In brain simulation, the states inside neurons will be updated at each time step, according to the integral of its dynamics equations from modeling methods. In this paper, the leaky-integrated and fire (LIF) model\cite{neural_dynamics} is used as the modeling method of neurons, with ordinary differential equations shown at the following:\\
\begin{equation}
\tau \frac{du_i(t)}{dt}=-u_i(t)+u_{rest}+R(I_{syn,i}(t) + I_{ext,i}(t))
\end{equation}
\begin{equation}
\rm{if} \, \mathit{u(t)}=\theta \Longrightarrow \mathit{u \rightarrow u_r \quad \rm{within} \quad \mathit{(t, t+t_{RP}]}}
\end{equation}
where $u_i$ is membrane potential, $u_{rest}$ is the resting potential, $\tau$ is membrane time constant, $R$ is membrane resistance, $I_{syn,i}$ is synaptic current, $I_{ext,i}$ is external current, $t$ is time, $\theta$ is spike threshold, $u_r$ is reset value of membrane potential, and $t_{RP}$ is refractory period. As shown in Fig.~\ref{fig:neuron_spike}, membrane potential will rapidly rise and then fall into $u_r$ during $(t, t+t_{RP}]$, which is called a spike. The synaptic current\cite{neural_dynamics} can be described as:\\
\begin{equation}
I_{syn,i}(t)=\sum_j\sum_f\delta(t-t^f_j)W_{ji}(t)g_{syn}(u_i(t)-E_{syn})
\end{equation}
where $W_{ji}$ is a synaptic weight from neuron $j$ to $i$, $g_{syn}$ is a time-dependent function of synaptic conductance, $t^f_j$ is the arrival time of spike $f$ from neuron $j$, and $E_{syn}$ is reversal potential of a synaptic ion channel. An exponential function was used as $g_{syn}$.\\
Different from other simulators, most ideas of optimizations in CORTEX are not from practical coding or test, but from graph abstraction of SNNs with theoretical traceable analysis, guiding the parallelization from the process level to the thread level. In the section of OVERVIEW OF CORTEX, graph abstraction of SNNs will be described in detail.\\
\subsection{Sparsity in Spatial and Temporal Domains}
In brain simulation, the computing intensity is much lower than that in primary HPC applications, because of its sparsity in both spatial and temporal domains, which presents the main challenges. First, it's easy to realize that the synaptic interactions inside a brain can be really sparse, with only 160 trillion synapses for 86 billion neurons in our human brain. In other words, each neuron only connects with thousands of others, which are much fewer than the neurons themselves. Therefore, the computation among these sparse interactions must be hard to be accelerated by mainstream computing architectures designed with increasingly higher ratios of FLOPS$/$Byte. Second, another type of sparsity is shown in temporal domain, related to varied synaptic delays. Different from other graph computations, synaptic interactions from pre-synaptic neurons to post-synaptic neurons will not take effect immediately but after a few time steps, where the delays of different synaptic interactions can be not exactly the same. In general, these varied synaptic delays must be the main reason for low computing efficiency.\\
\subsection{Ratio of Computation and Communication}
As a typical sparse problem, the ratio of computation and communication has a significant impact on general performance. Also, we are keen to discuss the reason behind the chosen LIF neuron model\cite{iaf_psc_exp_1, iaf_psc_exp_2, iaf_psc_alpha}. In the section on VERIFICATION AND EVALUATION, the selected modeling method has been generally applied in brain science investigation\cite{multi_area_1, multi_area_2, multi_area_3, potjans_2014}, because of its lightweight computing intensity, but precise approximation in neural dynamics, which is only slightly different from Hodgkin–Huxley neuron model\cite{hh_neuron} incorporating more biological details with much higher computing intensity.\\
Simulations\cite{fdu_brain} using neuron models with very high computing intensity show absolutely better results in scalability, which can be regarded as "good" cases for massive parallel computing. In our consideration, these "good" cases exhibits only the upper bound performance of the simulator, which is too trivial to demonstrate the contribution in optimizations. Therefore, only "bad" cases, with low ratio of computation to communication, are driving the design and implementation of brain simulation towards higher actual performance with maximal problem size, facing the sparsity of brain architecture in reality.\\
\subsection{Current State of The Art}
Various types of large-scale brain simulation have been proposed with levels of description and scales according to their purposes. A large-scale simulation of the rodent primary somatosensory cortex, with detailed morphological features of neurons, has been reported by the Blue Brain Project\cite{Markram_Cell}. In this case, the study focused on conditions of synchronization and influences of morphological features on network activities, simulated on the supercomputer Piz Daint at EPFL. Large-scale simulations of the cerebral cortex have been performed by various research teams. On Blue Gene P\cite{blue_gene_p}, a cat-scale corticothalamic circuit with 1.6 billion neurons was presented by IBM, using the C2 simulator. Subsequently, using the NEST Simulator \cite{kunkel_2012, kunkel_2014}, Kunkel and her colleagues performed non-spatial cortical models with 1.8 billion neurons and ten trillion of synapses on the K computer. Forked from NEST Simulator, MONET\cite{igarashi_monet} has successfully simulated a layered sheet-type of a model cerebral cortex with five billion neurons on the K computer by Igarashi and his colleagues, also a human-scale cerebellar model with 68 billion neurons\cite{yamazaki_review}. However, the cases we mentioned above shouldn't be a direct benchmark or baseline, because there might be some tricks in modeling methods or accuracy compression on floating point numbers with single or half precision. Up till now, human-scale whole brain simulation with cerebral cortex, cerebellum, and thalamus has never been conducted to date. Above all, NEST Simulator is the SOTA\cite{snn_robot_mouse} available with open-source code. Therefore, the comparison will be shown between CORTEX and NEST Simulator, in the section of VERIFICATION AND EVALUATION.\\
\subsection{System and Environment of Fugaku}
The supercomputer Fugaku\cite{fugaku} is a homogeneous, CPU-based distributed machine with 158,976 nodes. Based on the ARM v8.2A architecture, the CPU of the system, A64FX, has 32 GB HBM2 memory and 48 compute cores, with total memory bandwidth 1024 GB/s. 12 compute cores form one Core Memory Group (CMG). On the ALUs side, there is a couple of 512-bit wide SIMD units inside each compute core, where the whole CPU performs 3.072 TF/s for double-precision, 6.144 TF/s for single-precision, and 12.288 TF/s for half-precision. As to cache, Each compute core contains 64 KB of L1 data cache, and 32 MB L2 cache is shared by all. For interconnect, low latency and high throughput are achieved by a 6-dimensional mesh/torus network, with a 6.8 GB/s link bandwidth and 40.8 GB/s of injection bandwidth per compute node.\\
Using a hybrid parallelized strategy, an MPI process is assigned to each CMG, with an OpenMP thread allocated to each compute core, to achieve maximal bandwidth and minimal latency in this NUMA architecture. On the compiler side, the Fujitsu C/C++ compiler 4.10 is used in Trad mode with automatic vectorization and software pipelining.\\
\section{OVERVIEW OF CORTEX}
In this paper, CORTEX, our highly optimized framework for brain simulation on Fugaku supercomputer is proposed, focusing on biological data driven brain architecture, based on the graph abstraction. In this section, the concepts and principles will be described first, showing the basic ideas from which we start.\\
\subsection{Graph Abstraction of SNNs}
\begin{figure}[h!] 
	\centering
    \begin{subfigure}{0.2\textwidth}
        \centering
        \includegraphics[width=0.6\textwidth]{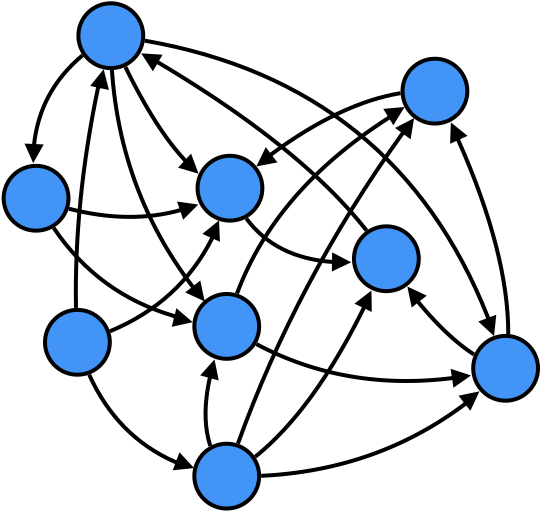}
        \caption{Graph Abstraction}
        \label{fig:graph_abstraction}
    \end{subfigure}
    \begin{subfigure}{0.2\textwidth}
        \centering
        \includegraphics[width=0.8\textwidth]{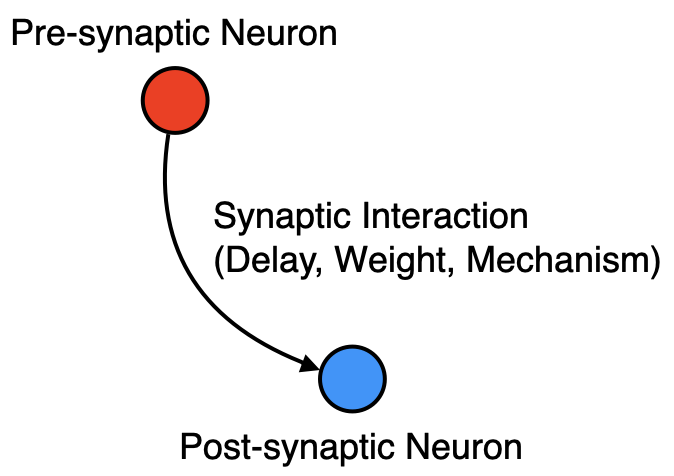}
        \caption{Basic Elements}
        \label{fig:basic_elements}
    \end{subfigure}
    \caption{Graph Abstraction of SNNs}
    \label{fig:grahp_snn}
\end{figure}
In a graph to represent the brain architecture, each vertex can be a neuron, and each edge is a synaptic interaction from, or input to, a neuron (vertex). The neural dynamics are computed on neurons, where synaptic interactions act on target neurons along edges. These basic abstractions of SNNs are shown in Fig.~\ref{fig:grahp_snn}, and we are going to describe how they form a brain architecture in simulation.\\
\subsubsection{Indegree and Outdegree Sub-graph}
\begin{figure}[h!] 
	\centering
    \begin{subfigure}{0.2\textwidth}
        \centering
        \includegraphics[width=0.6\textwidth]{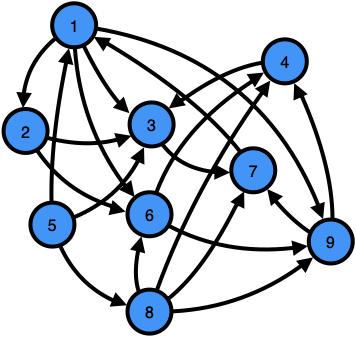}
        \caption{$G=(V,E)$}
        \label{fig:g=(v,e)}
    \end{subfigure}
    \begin{subfigure}{0.2\textwidth}
        \centering
        \includegraphics[width=0.6\textwidth]{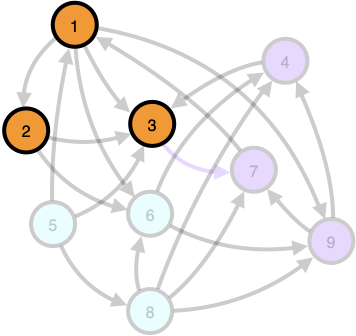}
        \caption{$\tilde{V} \subseteq V$}
        \label{fig:v_v}
    \end{subfigure}
    \begin{subfigure}{0.2\textwidth}
        \centering
        \includegraphics[width=0.6\textwidth]{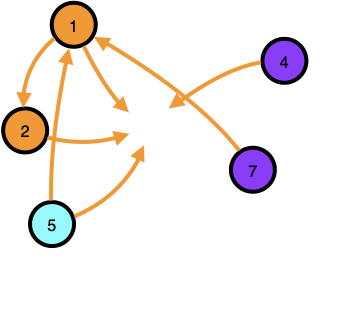}
        \caption{Pre-Vertices and Edges of the Indegree Sub-graph}
        \label{fig:pre_e_indegree}
    \end{subfigure}
    \begin{subfigure}{0.2\textwidth}
        \centering
        \includegraphics[width=0.6\textwidth]{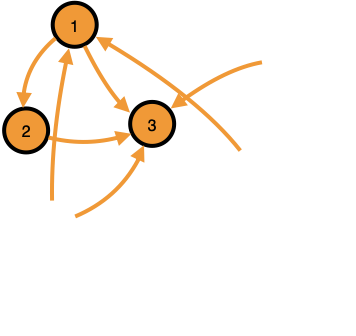}
        \caption{Post-Vertices and Edges of the Indegree Sub-graph}
        \label{fig:post_e_indegree}
    \end{subfigure}
    \begin{subfigure}{0.2\textwidth}
        \centering
        \includegraphics[width=0.6\textwidth]{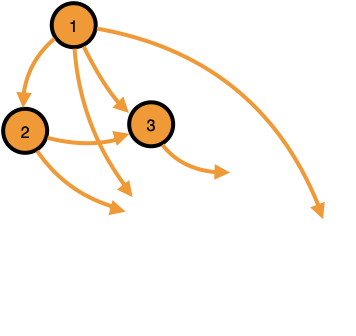}
        \caption{Pre-Vertices and Edges of the Outdegree Sub-graph}
        \label{fig:pre_e_outdegree}
    \end{subfigure}
    \begin{subfigure}{0.2\textwidth}
        \centering
        \includegraphics[width=0.6\textwidth]{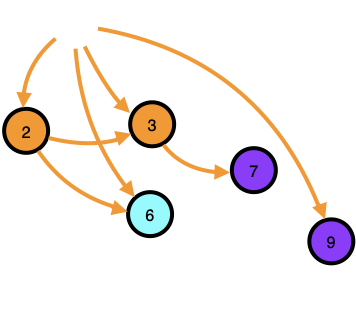}
        \caption{Post-Vertices and Edges of the Outdegree Sub-graph}
        \label{fig:post_e_outdegree}
    \end{subfigure}
    \caption{A Sub-graph of Indegree and Outdegree Format}
    \label{fig:indegree_outdegree_subgraph}
\end{figure}

A directed graph is an ordered pair $G=(V,E)$, comprising:
\begin{itemize}
  \item {$V$, a set of vertices, which can be regarded as neurons;}
  \item {$E\subseteq \{(x, y)| (x,y)\in V^2 $ and $ x \ne y\}$, a set of edges, as synaptic interactions, which are ordered pairs of vertices, where the condition: $ x \ne y$ can be ignored in some SNNs.}
\end{itemize}
Also, there can be a triplet:
\begin{equation}
^*\!S=(^*\!V^{pre},^*\!V^{post},^*\!E)
\end{equation}
where $* \in \{in, out\}$, defined as an indegree or outdegree format of graph $G$, comprising:
\begin{itemize}
  \item {$^*\!V^{pre} \subseteq V$, the set of pre-vertices;}
  \item {$^*\!V^{post} \subseteq V$, the set of post-vertices.}
  \item {$^*\!E\subseteq \{(x, y)| x \in V^{pre} $ and $ y \in V^{post}\}$, the set of edges, with a bijection $f:E \rightarrow ^*\!E$}
\end{itemize}
However, if the scale of simulation becomes large enough, the entire graph should be stored in distributed memory with domain decomposition. Assumed that $\tilde{V}$ is a subset of $V$, the indegree sub-graph based on $\tilde{V}$ can be well defined as:
\begin{equation}
^{in}\!S(\tilde{V})=(^{in}\!\tilde{V}^{pre},\tilde{V},^{in}\!\tilde{E})
\end{equation}
where $^{in}\!\tilde{V}^{pre} = \{x| (x,y) \in ^{in}\!\tilde{E}\}$, $^{in}\!\tilde{E} = \{(x,y)| (x,y) \in ^{in}\!E$ and $y \in \tilde{V}\}$.\\
Similarly, the outdegree sub-graph should be:
\begin{equation}
^{out}\!S(\tilde{V})=(\tilde{V},^{out}\!\tilde{V}^{post},^{out}\!\tilde{E})
\end{equation}
with $^{out}\!\tilde{V}^{post} = \{y| (x,y) \in ^{out}\!\tilde{E}\}$, $^{out}\!\tilde{E} = \{(x,y)| (x,y) \in ^{out}\!E$ and $x \in \tilde{V}\}$.\\
    When $^{*}\!S$ is defined without specification of $\tilde{V}$, definition (4) should be adopted. When $^{*}\!S(\tilde{V})$ has been specified with $\tilde{V}$, as a function from sets of vertices to sub-graphs, we use the definition in (5) or (6), without misunderstanding. Then, it is straightforward to define binary operation $\circledast$ on the following:
\begin{equation}
    ^*\!S_{a} \circledast ^*\!S_{b} = (^*\!V^{pre}_{a} \odot ^*\!V^{pre}_{b}, ^*\!V^{post}_{a} \odot ^*\!V^{post}_{b}, ^*\!E_{a} \odot ^*\!E_{b})
\end{equation}
where $(\circledast, \odot) \in \{(\barwedge, \cap), (\veebar, \cup)\}$. Obviously, each of them is commutative and associated, further, there is a homomorphism:
\begin{equation}
    ^*\!S(V_a) \circledast ^*\!S(V_b) = ^*\!S(V_a \odot V_b)
\end{equation}
An illustration can be shown as Fig.~\ref{fig:indegree_outdegree_subgraph}.
\subsubsection{Synaptic Interactions on Sub-graphs}
\begin{figure}[h!] 
	\centering
    \begin{subfigure}{0.2\textwidth}
        \centering
        \includegraphics[width=0.6\textwidth]{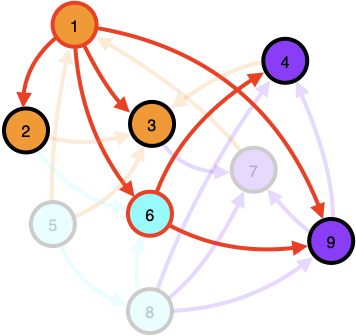}
        \caption{Indegree Spiking Graph}
        \label{fig:in_spk_graph}
    \end{subfigure}
    \begin{subfigure}{0.2\textwidth}
        \centering
        \includegraphics[width=0.6\textwidth]{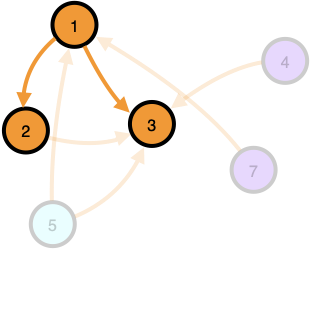}
        \caption{Indegree Sub-graph Tangerine}
        \label{fig:in_proc_tangerine}
    \end{subfigure}
    \begin{subfigure}{0.2\textwidth}
        \centering
        \includegraphics[width=0.6\textwidth]{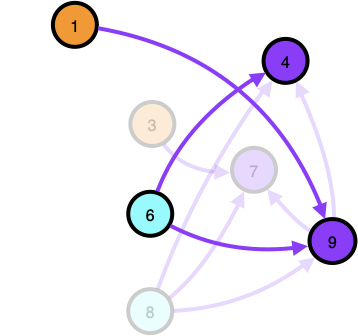}
        \caption{Indegree Sub-graph Grape}
        \label{fig:in_proc_grape}
    \end{subfigure}
    \begin{subfigure}{0.2\textwidth}
        \centering
        \includegraphics[width=0.6\textwidth]{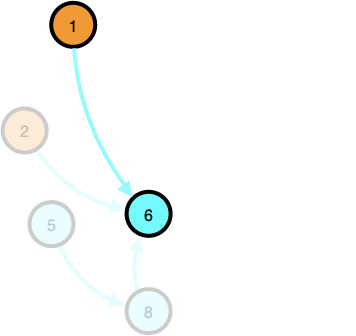}
        \caption{Indegree Sub-graph Ice}
        \label{fig:in_proc_ice}
    \end{subfigure}
    \caption{Synaptic Interactions on the Indegree Sub-graph: In Fig.\ref{fig:in_spk_graph}, spiking neurons with active synaptic interactions have been highlight with red border (No. 1 and No.6). At each time step, once the spiking graph has been well defined, synaptic interactions can take effect on corresponding post-synaptic neurons without dependency between different indegree sub-graphs.}
    \label{fig:syn_indegree}
\end{figure}

\begin{figure}[h!] 
	\centering
    \begin{subfigure}{0.2\textwidth}
        \centering
        \includegraphics[width=0.6\textwidth]{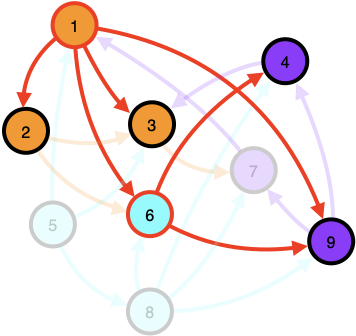}
        \caption{Outdegree Spiking Graph}
        \label{fig:out_spk_graph}
    \end{subfigure}
    \begin{subfigure}{0.2\textwidth}
        \centering
        \includegraphics[width=0.6\textwidth]{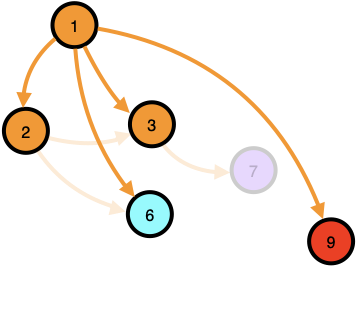}
        \caption{Outdegree Sub-graph Tangerine}
        \label{fig:in_proc_tangerine}
    \end{subfigure}
    \begin{subfigure}{0.2\textwidth}
        \centering
        \includegraphics[width=0.6\textwidth]{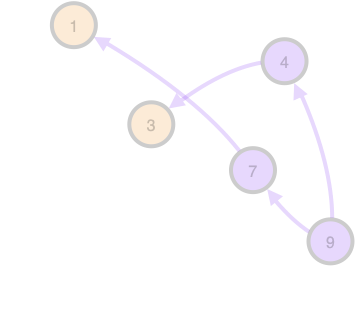}
        \caption{Outdegree Sub-graph Grape}
        \label{fig:in_proc_grape}
    \end{subfigure}
    \begin{subfigure}{0.2\textwidth}
        \centering
        \includegraphics[width=0.6\textwidth]{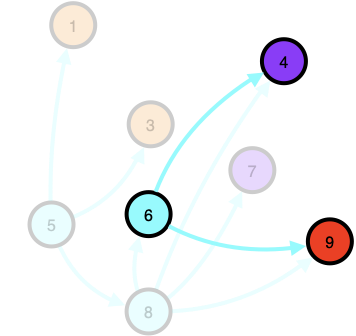}
        \caption{Outdegree Sub-graph Ice}
        \label{fig:in_proc_ice}
    \end{subfigure}
    \caption{Synaptic Interactions on the Outdegree Sub-graph: In Fig.~\ref{fig:out_spk_graph}, spiking neurons with active synaptic interactions have been highlight with red border (No. 1 and No.6). At this time step, there are 2 synaptic interactions taking effect on neuron 9 in 2 outdegree sub-graphs respectively. Between each synaptic interaction with its nonlinear dynamics, the states of all post-synaptic neurons with the same ID should be synchronized among all sub-graphs.}
    \label{fig:syn_outdegree}
\end{figure}

Considering synaptic interactions, the question comes to which kind of sub-graphs we should choose for parallel implementation. First, there can be a well partition of $V$, into $n$ parts:
\begin{equation}
\{V_1, V_2, V_3,...,V_n\}
\end{equation}
where $V_i \cap V_j = \varnothing$ with $i \ne j$, and $\bigcup^{n}_1 V_i = V$. Also, there are $n$ sub-graphs as:
\begin{equation}
\{^*\!S(V_1),^*\!S(V_2),^*\!S(V_3),...,^*\!S(V_n)\}
\end{equation}
where each one is corresponded to one process.\\
In one time step, there are very few pre-synaptic neurons in spiking state, with their corresponding synaptic interactions and post-synaptic neurons. So, there can be a set of all spiking pre-synaptic neurons $^*\!V^{pre}_{s}$, and the corresponding post-synaptic neurons $^*\!V^{post}_{s}$ with synaptic interactions $^*\!E_{s}$, forming a spiking graph $^*\!S_{s} = (^*\!V^{pre}_{s}, ^*\!V^{post}_{s}, ^*\!E_{s})$. For a sub-graph $^*\!S(V_i)$ , its spiking sub-graph can be defined as:
\begin{equation}
    ^*\!S_{s}(V_i) = ^*\!S(V_i) \barwedge ^*\!S_{s}
\end{equation}
Then, as we mentioned before, the synaptic interactions will effect at post-synaptic neurons, where the data instances of them should be writable. In parallel implementation, we want that all write operations can be performed without any dependency between spiking sub-graphs. Based on what we mentioned before, dependencies between 2 arbitrary sub-graphs can be expressed as:
\begin{equation}
^*\!S_{a} \barwedge ^*\!S_{b}
\end{equation}
Because of the homomorphism in (8), with the commutative and associated operation, we have:
\begin{equation}
^*\!S_{s}(V_i) \barwedge ^*\!S_{s}(V_j) = ^*\!S(V_i \cap V_j) \barwedge ^*\!S_{s}
\end{equation}
Obviously, for indegree sub-graphs, we have:
\begin{equation}
^{in}\!S(V_i \cap V_j) = (^{in}\!V^{pre}_i \cap ^{in}\!V^{pre}_j, \varnothing, \varnothing)
\end{equation}
while in outdegree sub-graphs:
\begin{equation}
^{out}\!S(V_i \cap V_j) = (\varnothing, ^{out}\!V^{post}_i \cap ^{out}\!V^{post}_j, \varnothing)
\end{equation}
What we mentioned in (14) or (15) can indicate which kind of data should be synchronized for each modification, as shown in Fig.~\ref{fig:syn_indegree} and~\ref{fig:syn_outdegree}. Therefore, indegree sub-graphs should be the only choice.\\
\subsection{Simulation Procedure}
\begin{figure}[h!] 
	\centering
    \begin{subfigure}{0.24\textwidth}
        \centering
        \includegraphics[width=0.5\textwidth]{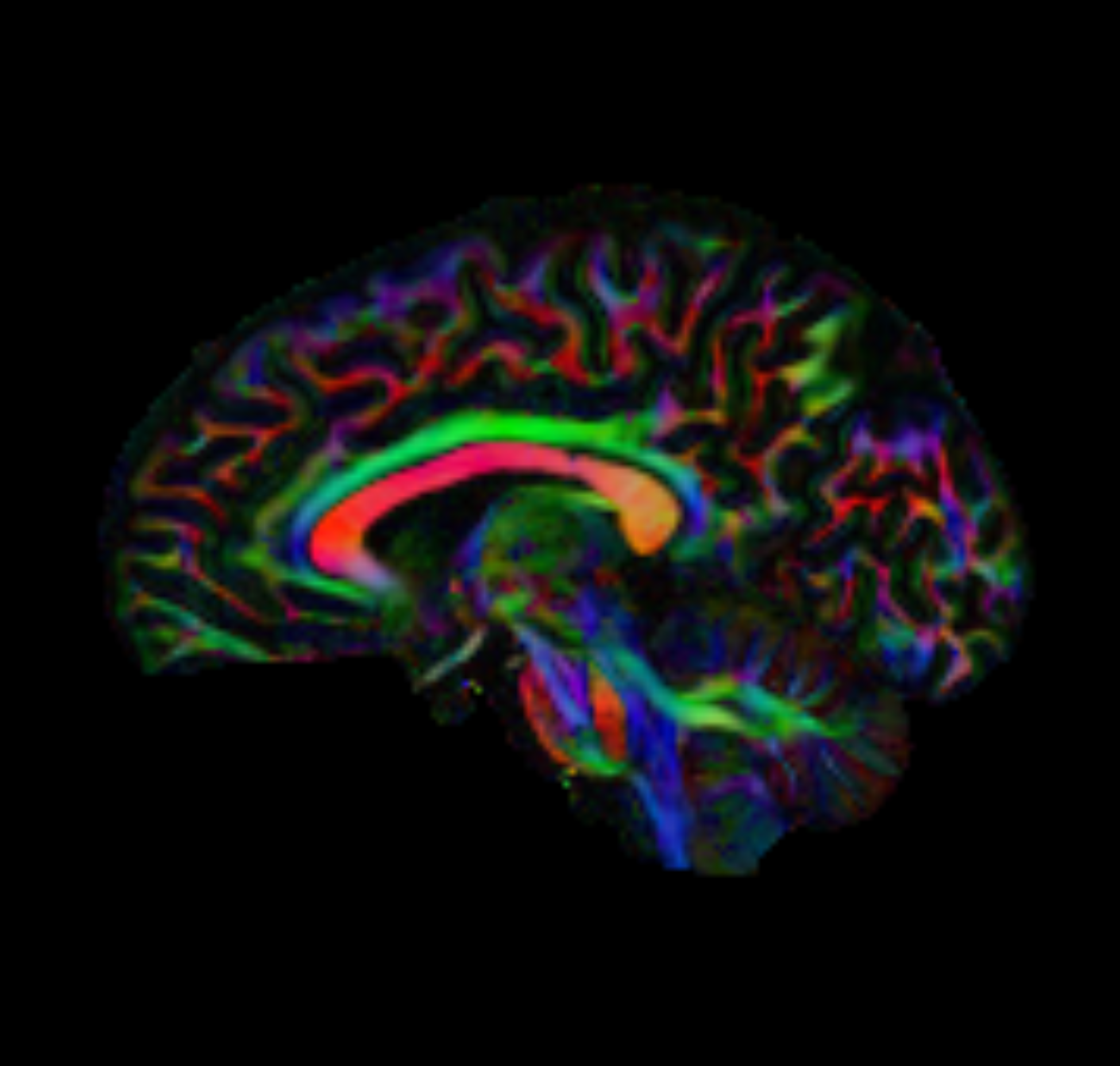}
        \caption{Connectome Data}
        \label{fig:brain_data}
    \end{subfigure}
    \begin{subfigure}{0.24\textwidth}
        \centering
        \includegraphics[width=0.5\textwidth]{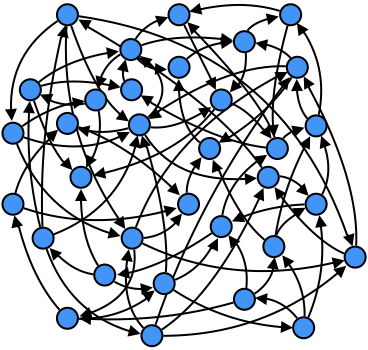}
        \caption{Problem Description}
        \label{fig:problem_description}
    \end{subfigure}
    \begin{subfigure}{0.24\textwidth}
        \centering
        \includegraphics[width=0.5\textwidth]{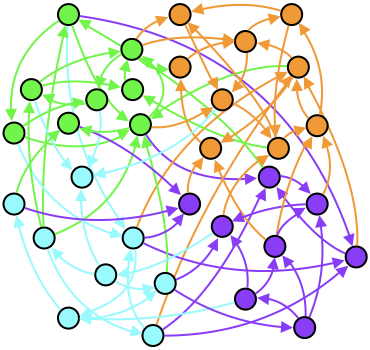}
        \caption{Domain Decomposition}
        \label{fig:domain_decomposition}
    \end{subfigure}
    \begin{subfigure}{0.24\textwidth}
        \centering
        \includegraphics[width=0.5\textwidth]{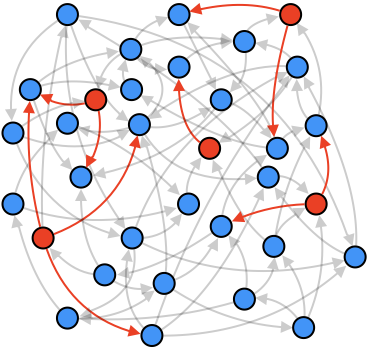}
        \caption{Simulation}
        \label{fig:simulation}
    \end{subfigure}
    \caption{Brain Simulation Procedure: fig (a) is an illustration of brain connectome data extracted from Diffusion magnetic resonance imaging. In fig (d), the neurons in color red are the spiking neurons, whose synaptic interactions with color red too will effect at the post-synaptic neurons after corresponding synaptic delay.}
    \label{fig:brain_simulation_procedure}
\end{figure}
As to run a brain simulation by CORTEX on Fugaku supercomputer, it might contain several steps. Fig.~\ref{fig:brain_simulation_procedure} describes 4 main steps of brain simulation. Different from other primary HPC applications like Computational Fluid Dynamics (CFD) or Molecular Dynamics (MD), only a few parts of synaptic interactions (edges) will be active in one time step. As shown in Fig.~\ref{fig:simulation}, the synaptic interactions from spiking pre-synaptic neurons in red are taking effect on their post-synaptic neurons. After that, all neurons (vertices) will update their variables like membrane potential by neural dynamics, generating new spiking pre-synaptic neurons, to start a new loop.\\
\section{MAJOR INNOVATIVE CONTRIBUTIONS}
Based on our graph abstraction of SNNs, the main idea for optimized parallelization has been clear. Then, it comes to implementation, facing the sparsity in both spatial and temporal domains. Efforts on storage, computing and networking are adopted for not only application-available problem size but also general performance of simulation. In brief, our key innovations are summarized as follows:\\
\begin{itemize}
  \item {A customized domain decomposition method comprising of 2 steps: Area-Processes Mapping and Multisection Division with Sampling Method, is introduced to achieve maximal problem size with high performance.}
  \item {A elaborate multi-threading parallel scheme without mutex or atomic operation when computing synaptic interactions on edges, which is the hotspot of whole simulation.}
  \item {An optimized spikes broadcast method with dedicated thread for communication, in order to overlap communication and computation as much as possible.}
\end{itemize}
\subsection{Domain Decomposition}
Fortunately, with the homomorphism in (8), the domain decomposition of graph into sub-graphs can be transferred to the partition on vertices. Then, the problem arises to generate appropriate partitions of vertices with their corresponding indegree sub-graphs to achieve best performance in simulation, which is a typical NP-hardness\cite{graph_np_hard}. Therefore, the solution should be generally derived using heuristics and approximation algorithms.\\
\subsubsection{Varied Density of Synaptic Interactions}
Before domain decomposition, the properties of biological data driven SNNs should be taken into consideration, owing to the varied density of both synaptic interactions and neurons.\\
\begin{figure}[h!] 
	\centering
    \includegraphics[width=0.25\textwidth]{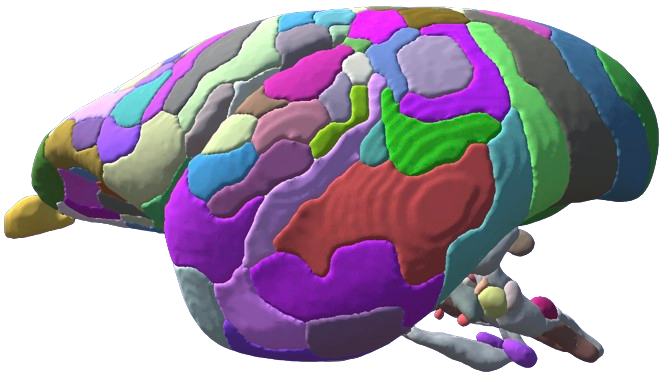}
    \caption{An example of marmoset brain atlas from Brain/MINDS Project\cite{brainatlas}, where each color patch represents a distinct brain area.}
    \label{fig:brain_atlas}
\end{figure}
As shown in Fig.~\ref{fig:brain_atlas}, with the Atlas, a brain architecture can be separated into areas. Each area is related to different macro behavior or brain function. For example, the primary vision cortex V1 area is related to vision and the primary motor cortex M1 area is related to motor control. As we can find from biological data, the density of synaptic interactions within one area is much higher than that between areas.\\
\subsubsection{Area-Processes Mapping}
\begin{figure}[h!] 
	\centering
    \begin{subfigure}{0.23\textwidth}
        \centering
        \includegraphics[width=0.8\textwidth]{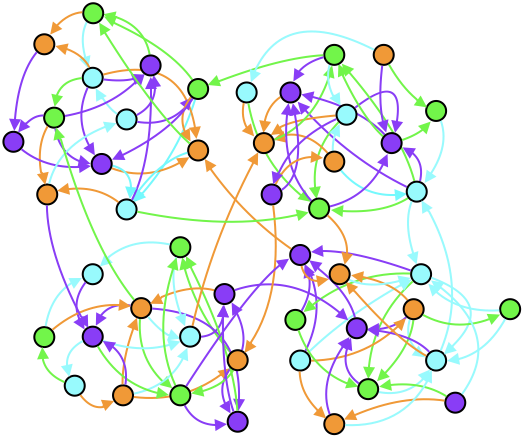}
        \caption{Random Equivalent Mapping}
        \label{fig:random_map}
    \end{subfigure}
    \begin{subfigure}{0.23\textwidth}
        \centering
        \includegraphics[width=0.8\textwidth]{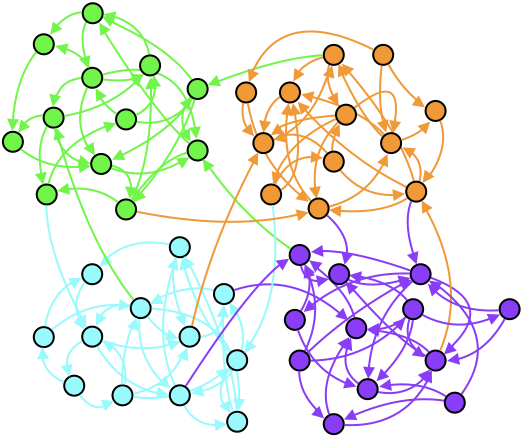}
        \caption{Area-processes Mapping}
        \label{fig:area_map}
    \end{subfigure}
    \caption{Mapping Methods for Domain Decomposition: Remote synaptic interactions with red is much less than local synaptic interactions with black.}
    \label{fig:map_method}
\end{figure}

\begin{figure}[h!] 
	\centering
    \begin{subfigure}{0.23\textwidth}
        \centering
        \includegraphics[width=0.8\textwidth]{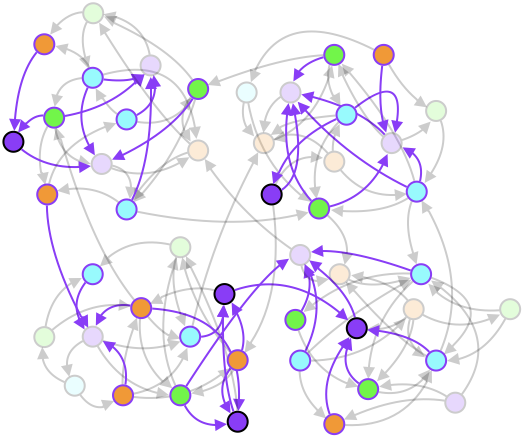}
        \caption{Pre-synaptic Neurons}
        \label{fig:random_pre_syn}
    \end{subfigure}
    \begin{subfigure}{0.23\textwidth}
        \centering
        \includegraphics[width=0.8\textwidth]{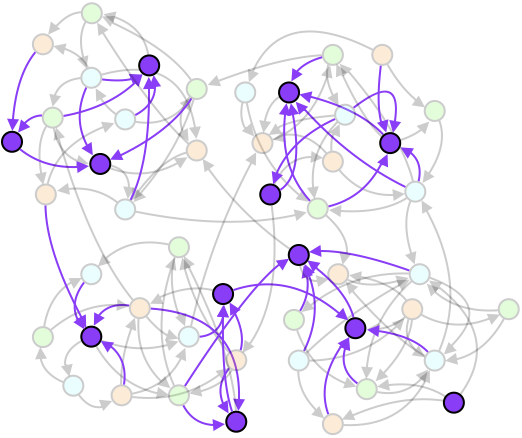}
        \caption{Post-synaptic Neurons}
        \label{fig:random_post_syn}
    \end{subfigure}
    \caption{Random Equivalent Mapping: There are 29 pre-synaptic neurons in process grape, while the number of post-synaptic neurons is 12, which is a quarter of total post-synaptic neurons.}
    \label{fig:random_pre_post}
\end{figure}

\begin{figure}[h!] 
	\centering
    \begin{subfigure}{0.23\textwidth}
        \centering
        \includegraphics[width=0.8\textwidth]{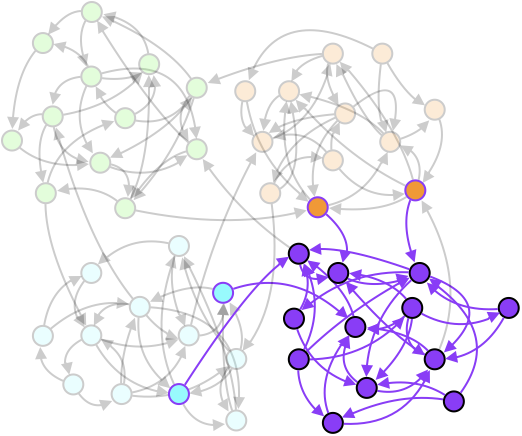}
        \caption{Pre-synaptic Neurons}
        \label{fig:map_pre_syn}
    \end{subfigure}
    \begin{subfigure}{0.23\textwidth}
        \centering
        \includegraphics[width=0.8\textwidth]{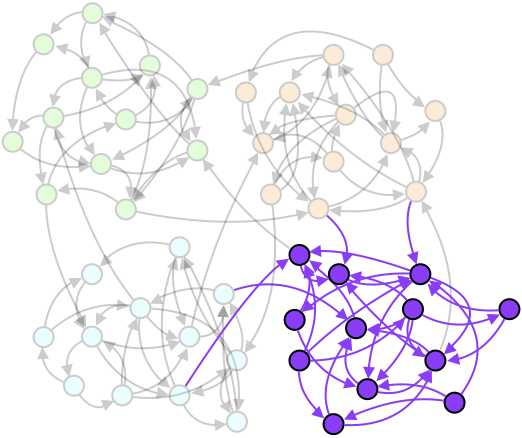}
        \caption{Post-synaptic Neurons}
        \label{fig:map_post_syn}
    \end{subfigure}
    \caption{Area-processes Mapping: There are 16 pre-synaptic neurons in process grape, which has been reduced obviously compared to Random Equivalent Mapping. And the number of post-synaptic neurons is 12, remaining unchanged.}
    \label{fig:map_pre_post}
\end{figure}

Intuitively, a principle for decomposition should be utilizing the areas described in the Atlas. First, all neurons are partitioned into $n$ parts as shown in (9), according to areas described in the Atlas. Then, indegree sub-graphs of each area can be well generated by the partition with (5). After that, memory consumption of each sub-graph can be estimated, making it easy to determine how many processes should be mapped to this area. Finally, as to the name, Area-Processes Mapping is to assign several specific processes to each area.\\
Additionally, compared with the naive approach: Random Equivalent Mapping as shown in~\ref{fig:random_map}, the advantage of Area-Processes Mapping is to reduce the number of pre-vertices (pre-synaptic neurons) in indegree sub-graphs. As shown in Fig.~\ref{fig:random_pre_post}, in Random Equivalent Mapping, the pre-vertices $^{in}\!V^{pre}_i$ of sub-graph $^{in}\!S(V_i)=(^{in}\!V^{pre}_i, V_i, ^{in}\!E_i)$ are randomly selected from $V$. In the worst condition, it can be $^{in}\!V^{pre}_i = V$, because the post-vertices in $V_i$ are possible to be pointed by edges from arbitrary pre-vertices in $V$. Then, regarding data instances, almost all pre-vertices in $V$ should be stored in each process. Therefore, in Random Equivalent Mapping, the memory consumption becomes unacceptable in large-scale simulation, with up to billions of neurons (vertices).\\
As shown in Fig.~\ref{fig:map_pre_post}, in Area-Processes Mapping, a sub-graph $S^{in}(V_i)$ can be further decomposed into $^{in}\!S^{l}(V_i)=(V_i, V_i, ^{in}\!E^{l}_i)$ and $^{in}\!S^{r}(V_i)=(^{in}\!V^{r}_i, V_i, ^{in}\!E^{r}_i)$, named local indegree sub-graph and remote indegree sub-graph respectively.
And we have:
\begin{equation}
^{in}\!S(V_i) = ^{in}\!S^{l}(V_i) \veebar ^{in}\!S^{r}(V_i) = (^{in}\!V_i \cup ^{in}\!V^{r}_i, V_i, ^{in}\!E^{l}_i \cup ^{in}\!E^{r}_i)
\end{equation}
where $V^{r}_i \cap V_i = \varnothing$, and so $E^{l}_i \cap E^{r}_i = \varnothing$.\\
As we have mentioned above, remote pre-synaptic neurons and remote synaptic interactions of each area are much less than the local ones as shown in Fig.~\ref{fig:area_map}, expressed as $n(^{in}\!V^{r}_i) \ll n(V_i)$ and $n(^{in}\!E^{r}_i) \ll n(^{in}\!E^{l}_i)$. In other words, most edges are in $^{in}\!S^{l}(V_i)$, whose number of post-vertices is fixed to $n(V_i)$. Obviously, in large-scale simulations, $n(V_i) \ll n(V)$, and we still get $n(V_i) + n(^{in}\!V^{r}_i) \ll n(V)$.\\
Therefore, with Area-Processes Mapping, the memory consumption of pre-synaptic neurons and post-synaptic neurons can be kept in a low level, making more memory available to synaptic interactions, which consumes the most memory. Utilizing varied densities of synaptic interactions between and inside areas, we are advancing the available problem size to the next level.\\
\subsubsection{Multisection Division with Sampling Method}
\begin{figure}[h!] 
	\centering
    \begin{subfigure}{0.24\textwidth}
        \centering
        \includegraphics[width=0.5\textwidth]{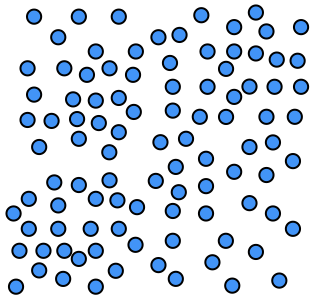}
        \caption{Original Distribution}
        \label{fig:org_distr}
    \end{subfigure}
    \begin{subfigure}{0.24\textwidth}
        \centering
        \includegraphics[width=0.5\textwidth]{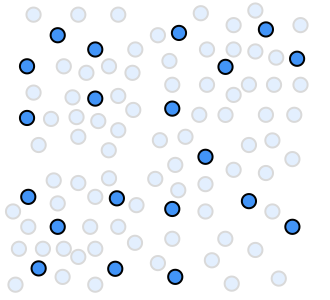}
        \caption{Sample Distribution}
        \label{fig:sample_distr}
    \end{subfigure}
    \begin{subfigure}{0.24\textwidth}
        \centering
        \includegraphics[width=0.5\textwidth]{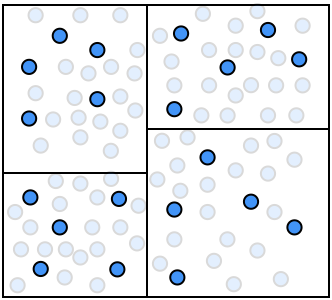}
        \caption{Division Generation}
        \label{fig:division_sample}
    \end{subfigure}
    \begin{subfigure}{0.24\textwidth}
        \centering
        \includegraphics[width=0.5\textwidth]{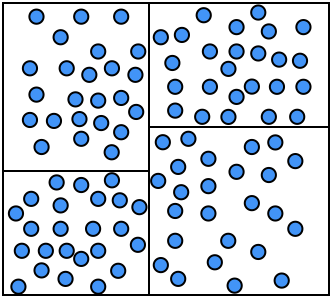}
        \caption{Division Application}
        \label{fig:division_app}
    \end{subfigure}
    \caption{Multisection Division with Sampling Method}
    \label{fig:MDSM}
\end{figure}

Area-Processes mapping is our first step in domain decomposition, because
each area might contain millions of neurons and billions of synaptic interactions, which are still hard to be stored in one process. Fortunately, in the definition of indegree sub-graphs, edges are bound to post-synaptic neurons, making it possible to perform a decomposition on post-synaptic neurons only to generate more indegree sub-graphs. In CORTEX, a stable implementation from FDPS\cite{fdps_1, fdps_2}, called Multisection Division with Sampling Method\cite{fdps_3}, is introduced. This method can generate divisions in grid cells for a roughly equal number of points (post-synaptic neurons) with their coordinates in euclidean space, even in a non-uniform distribution. After the post-synaptic neurons has been well defined by the generated divisions, an indegree sub-graph can be generated. As shown in Fig.~\ref{fig:MDSM}, there are several steps to generate such divisions and apply them into the original distribution of post-synaptic neurons with load balance.\\
\subsubsection{Load Balance}
From we mentioned above, the memory consumption of each process can be $O(n_{pre} + n_{post} + n_{edges})$, where edges consume most of the memory. From homomorphism in (8), the partition of edges can be transferred to the partition on vertices, which is easy to achieve load balance by 2 steps we mentioned before. First, in Area-Processes Mapping, memory consumption of each area in the adopted Atlas can be estimated. Then, each area will be mapped to several processes according to the estimation. After that, each area will be further decomposed into several sub-graphs by Multisection Division with Sampling Method, assuming that all properties are homogeneous inside an area.\\
\subsection{Multi-Threading Parallelization}
\begin{figure*}[t!] 
	\centering
    \begin{subfigure}{0.3\textwidth}
        \centering
        \includegraphics[width=0.8\textwidth]{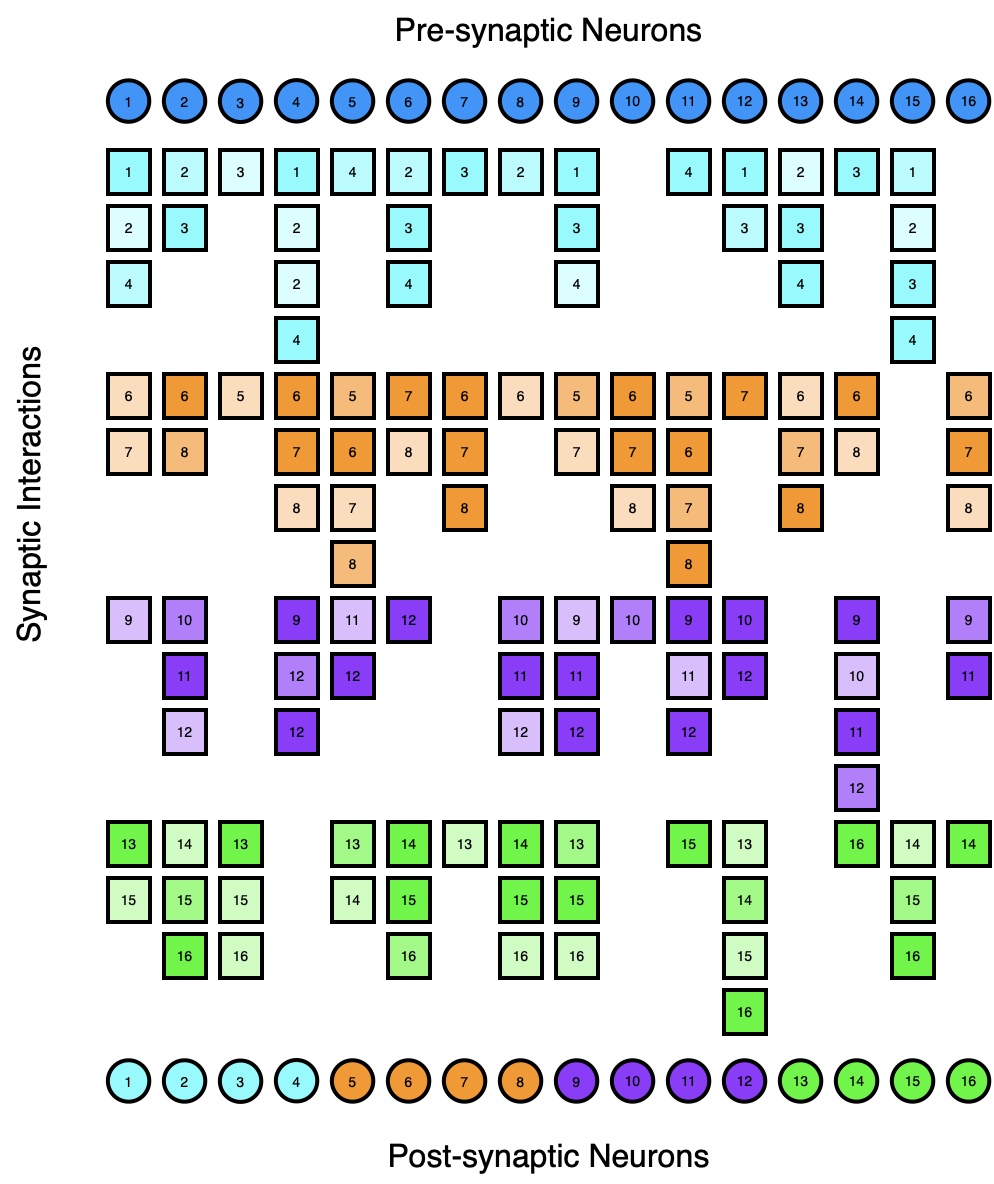}
        \caption{Increasing Index Order}
        \label{fig:org_syn_order}
    \end{subfigure}
    \begin{subfigure}{0.3\textwidth}
        \centering
        \includegraphics[width=0.8\textwidth]{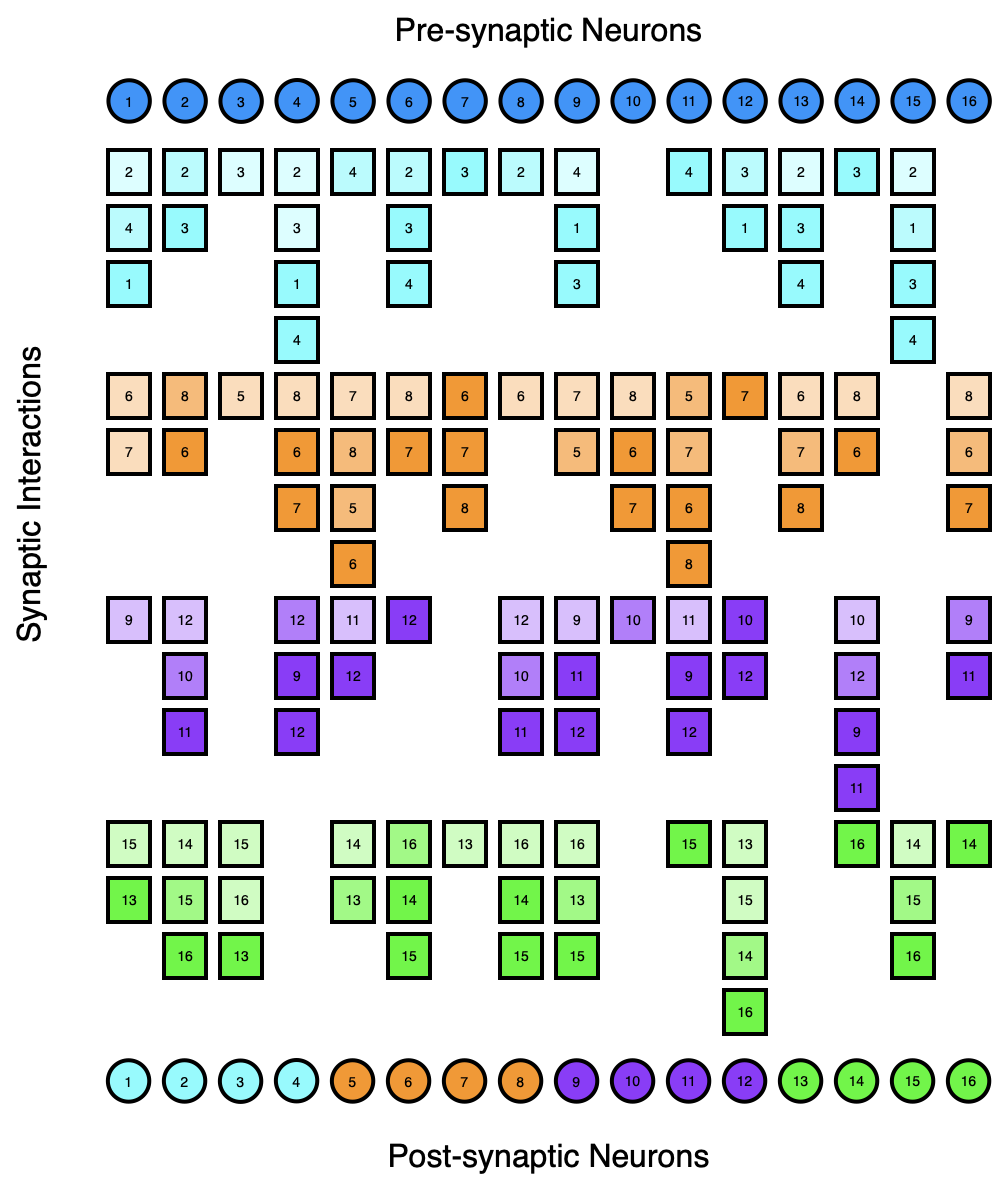}
        \caption{Increasing Delay Order}
        \label{fig:delay_syn_order}
    \end{subfigure}
    \begin{subfigure}{0.3\textwidth}
        \centering
        \includegraphics[width=0.8\textwidth]{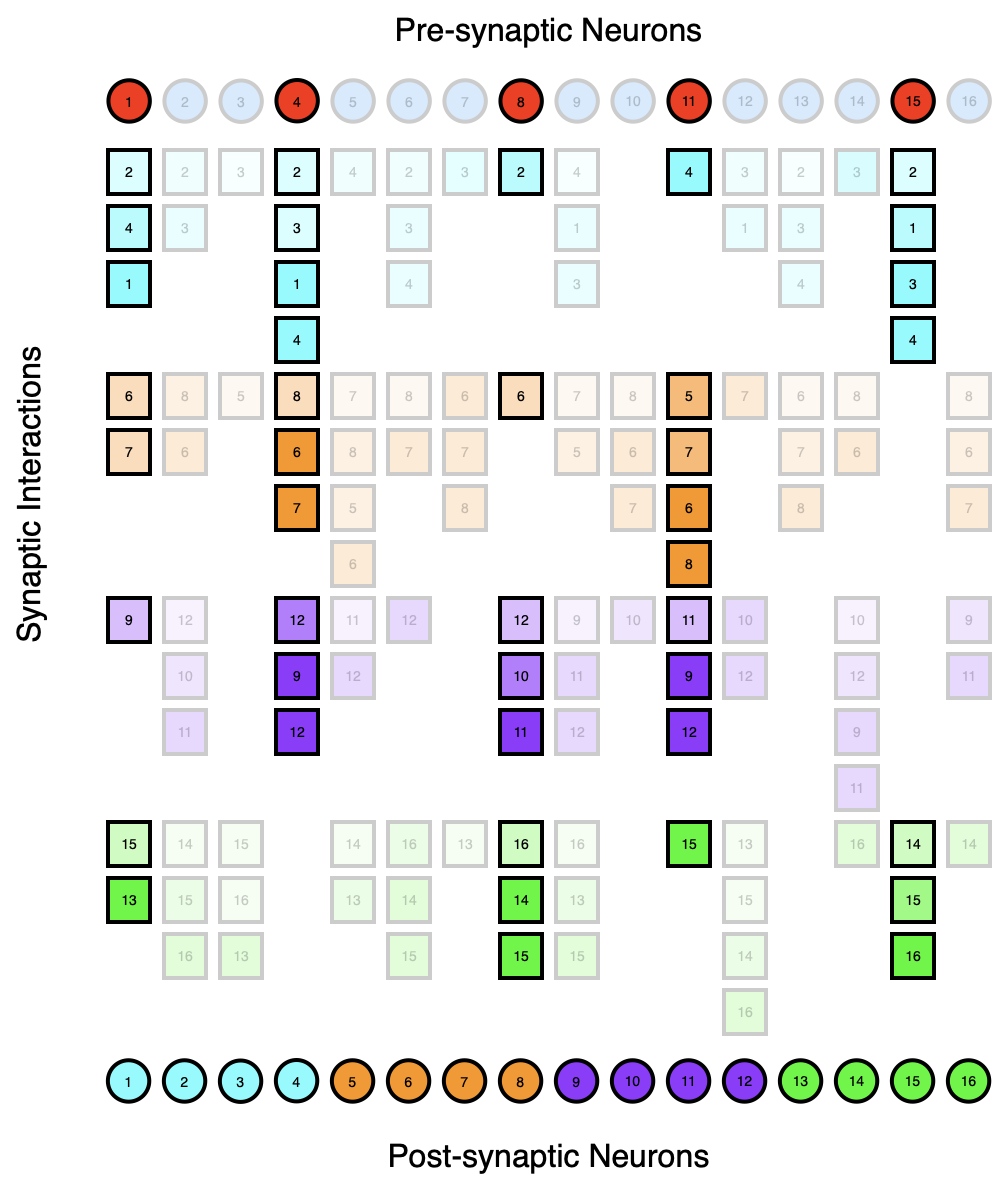}
        \caption{Spiking Sub-graph}
        \label{fig:spk_subgraph}
    \end{subfigure}
    \caption{Data Instance of Synaptic Interactions: Deeper in color means higher delay of this synaptic interaction.}
    \label{fig:data_ins}
\end{figure*}
In general, variables in synaptic interactions and post-synaptic neurons should be able to be read and written, while pre-synaptic neurons can be read-only and shared by all threads. Therefore, if a synaptic interaction and its post-synaptic neuron can be accessed by different threads, mutexes or atomic operations\cite{thu_snn_1, thu_snn_2} must be introduced, in order to avoid data racing, resulting in poor computing throughput. To transcend these barriers, an elaborate parallelization scheme has been implemented, which can handle massive processing of synaptic interactions at the thread level, without any mutex or atomic operation.\\
\subsubsection{Avoid Data Racing}
\begin{figure}[h!] 
	\centering
    \begin{subfigure}{0.2\textwidth}
        \centering
        \includegraphics[width=0.8\textwidth]{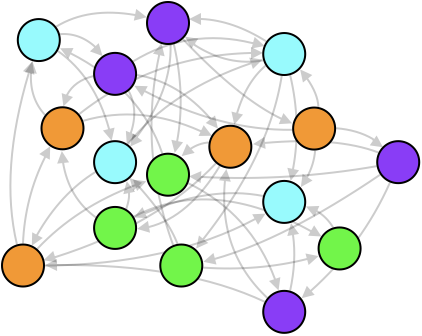}
        \caption{Vertices Mapping}
        \label{fig:v_map}
    \end{subfigure}
    \begin{subfigure}{0.2\textwidth}
        \centering
        \includegraphics[width=0.8\textwidth]{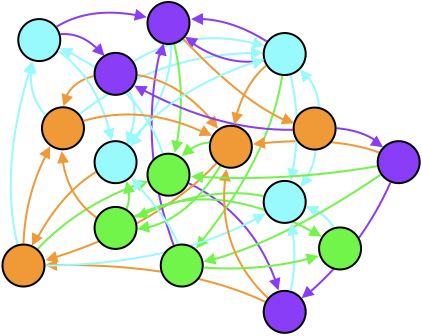}
        \caption{Edges Mapping}
        \label{fig:e_map}
    \end{subfigure}
    \caption{Vertices and Edges Mapping at the Thread Level}
    \label{fig:v_e_map_thread}
\end{figure}

\begin{figure}[h!] 
	\centering
    \begin{subfigure}{0.2\textwidth}
        \centering
        \includegraphics[width=0.8\textwidth]{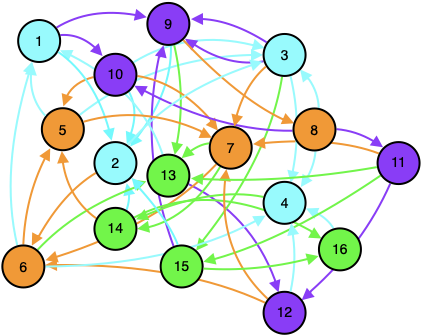}
        \caption{Directed Graph}
        \label{fig:g_format}
    \end{subfigure}
    \begin{subfigure}{0.2\textwidth}
        \centering
        \includegraphics[width=0.8\textwidth]{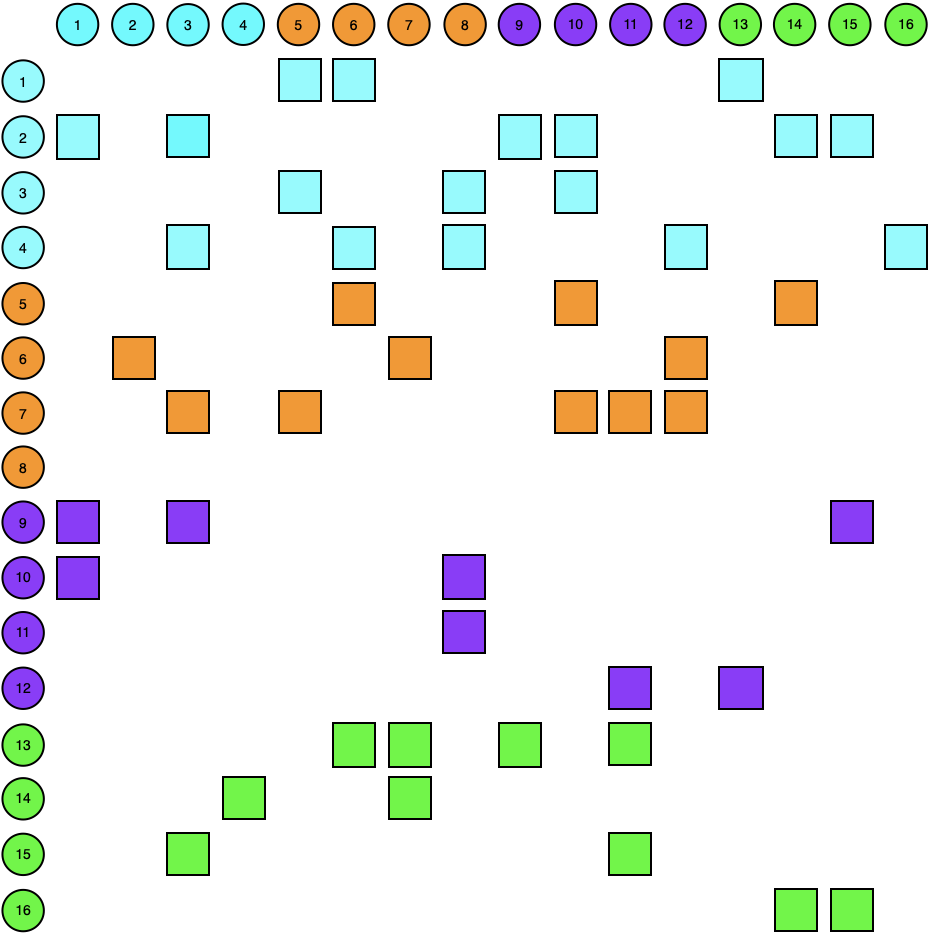}
        \caption{Sparse Matrix}
        \label{fig:m_format}
    \end{subfigure}
    \caption{Sparse Matrix of Synaptic Interactions}
    \label{fig:sp_m_syn}
\end{figure}

To avoid data racing, each synaptic interaction and its post-synaptic neuron should be accessed by a specific thread only. Therefore, the first thing is to partition the indegree sub-graph again and assign each synaptic interaction and its post-synaptic neuron to one thread in correspondence. Based on homomorphism in (8) with (9) and (10), sub-graph generations can be implemented by dividing all post-synaptic neurons with roughly equal numbers. After that, we map each vertex and edge to a specific thread as shown in Fig.~\ref{fig:v_e_map_thread}.\\
Intuitively, Fig.~\ref{fig:m_format} is an equivalent representation of Fig.~\ref{fig:g_format} in sparse matrix, where the columns are pre-synaptic neurons, the rows are post-synaptic neurons and elements are synaptic interactions. Each elements in sparse matrix (edges in graph) is assigned to a specific thread, as shown by its color, which is the same thread assigned to its post-synaptic neurons. In computation, each thread will only access to the synaptic interactions and post-synaptic neurons assigned with itself, where pre-synaptic neurons can be shared by all threads, because they are read-only during the computation of synaptic interactions, as we mentioned above.\\
As shown in Fig.~\ref{fig:org_syn_order}, there is the data instance of a sub-graph in memory, including pre-synaptic neurons, synaptic interactions and post-synaptic neurons. Except color 'aqua' for shared accessing to all threads, each color in each element represents a specific thread with accessing permission to this element. For synaptic interactions, deeper in color means higher delay, and the number is the pointer to specific post-synaptic neuron. Another reason for using such data format is that a spiking sub-graph of (11) can be easily generated. Because the only necessary information is the spiking pre-synaptic neurons, the corresponding edges and post-synaptic neurons can be conveniently found, as shown in Fig.~\ref{fig:spk_subgraph}.\\
\subsubsection{Varied Synaptic Delays}
\begin{figure*}[t!] 
	\centering
    \begin{subfigure}{0.3\textwidth}
        \centering
        \includegraphics[width=0.8\textwidth]{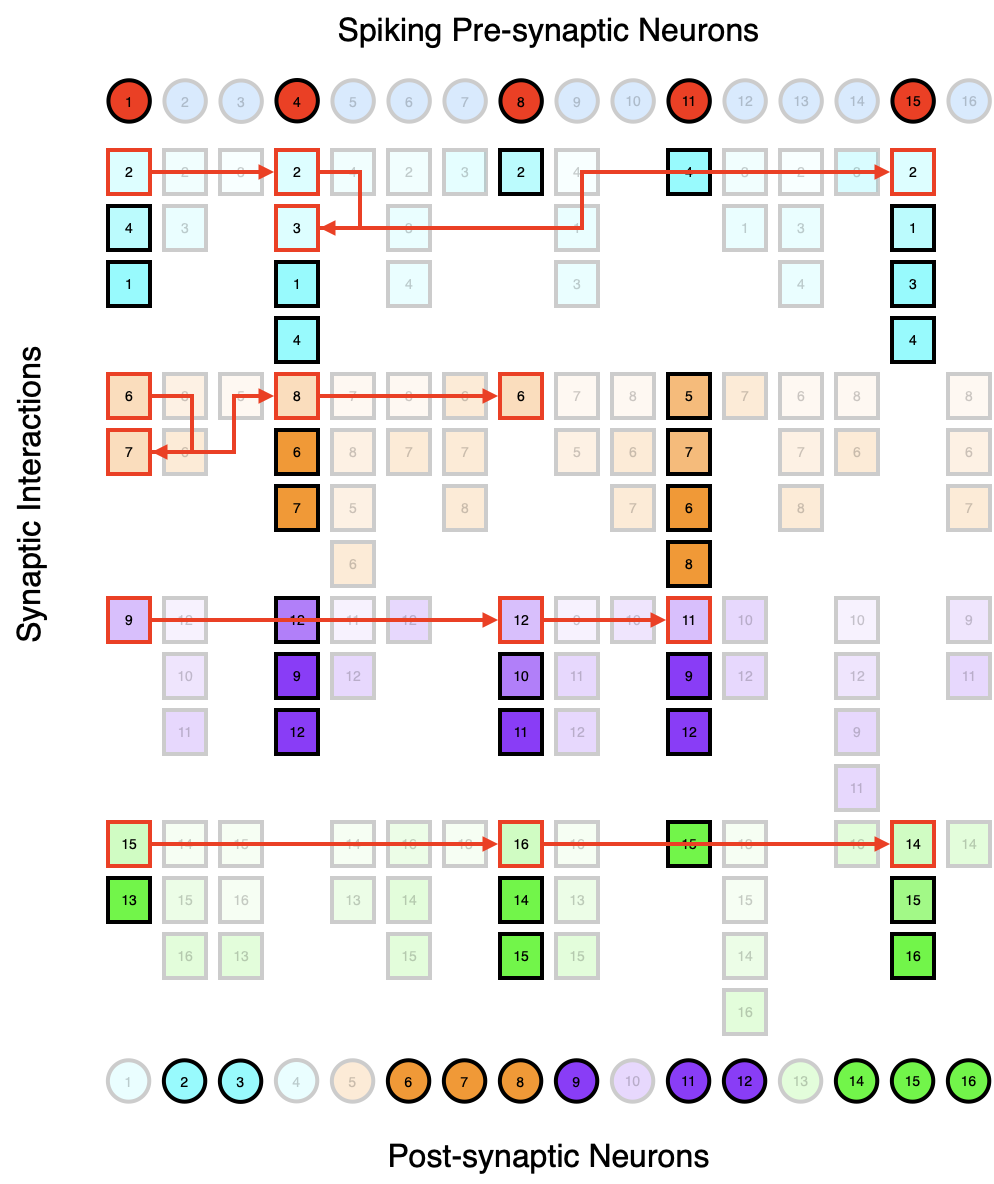}
        \caption{Delay Step 1}
        \label{fig:spk_comp_1}
    \end{subfigure}
    \begin{subfigure}{0.3\textwidth}
        \centering
        \includegraphics[width=0.8\textwidth]{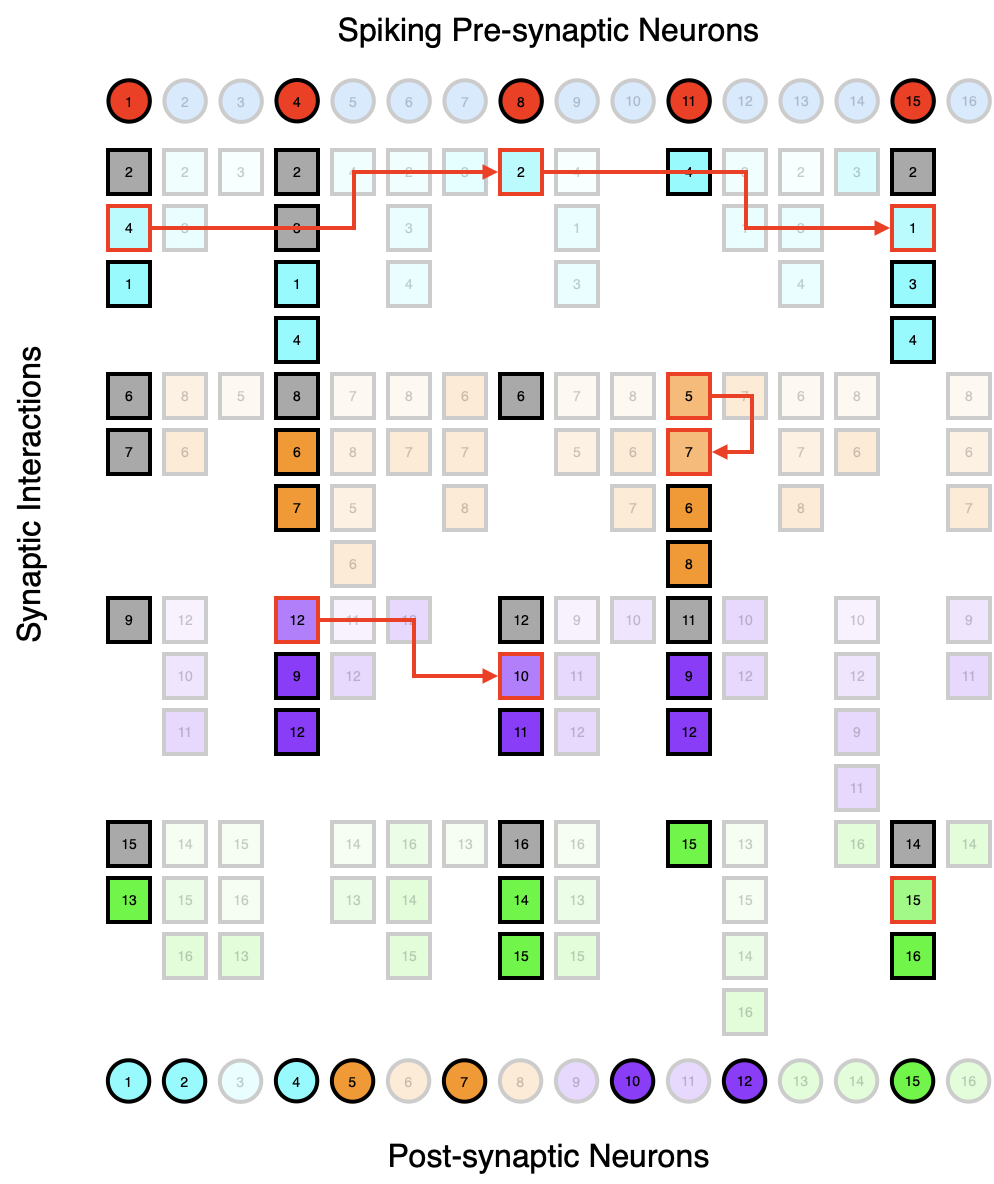}
        \caption{Delay Step 2}
        \label{fig:spk_comp_2}
    \end{subfigure}
    \begin{subfigure}{0.3\textwidth}
        \centering
        \includegraphics[width=0.8\textwidth]{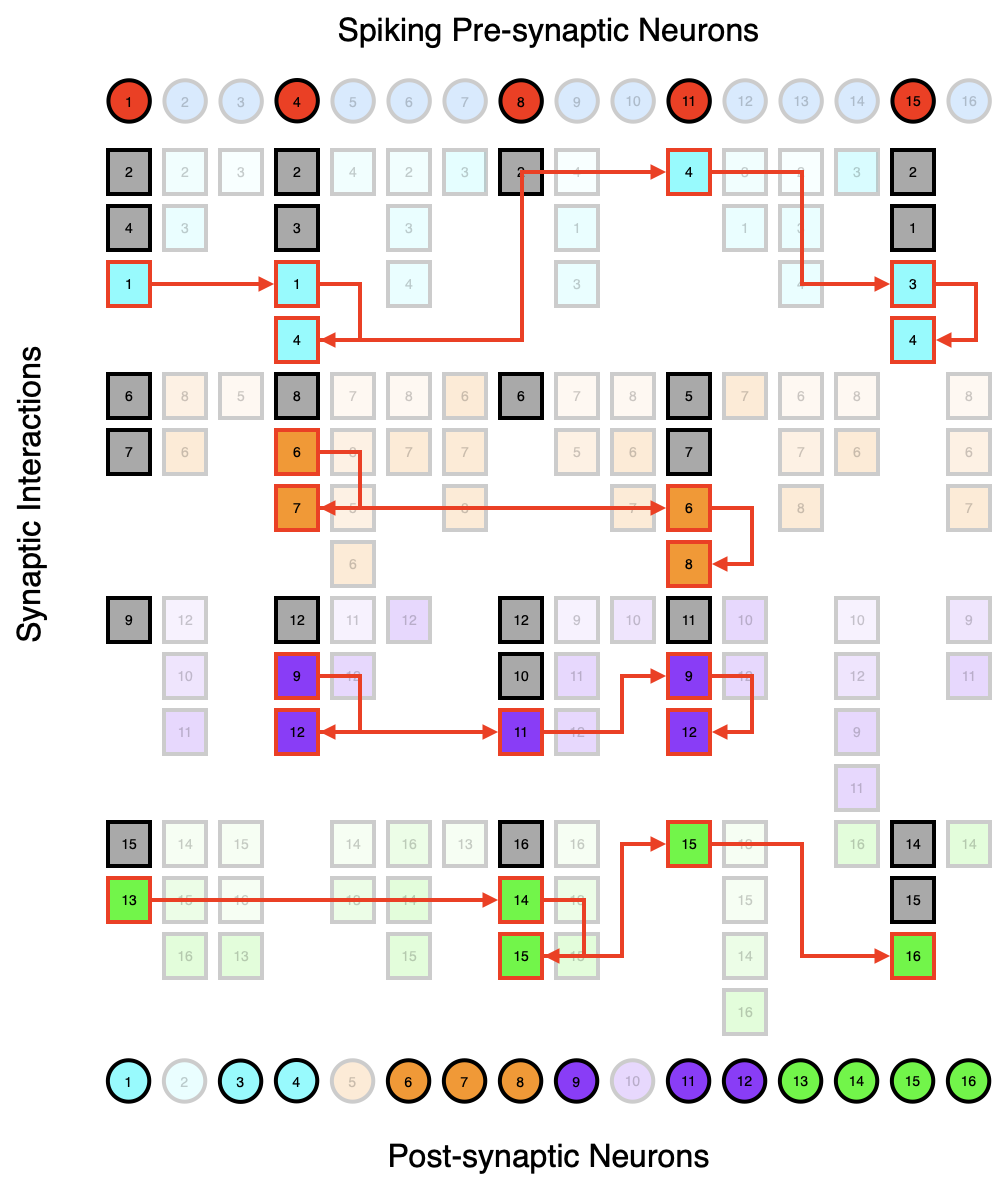}
        \caption{Delay Step 3}
        \label{fig:spk_comp_3}
    \end{subfigure}
    \caption{Synaptic Interactions with Varied Delays: The elements with red border are current accessing elements in this time step, and the elements in color 'aluminium' are the elements which have been accessed in the past}
    \label{fig:spk_delay_comp}
\end{figure*}

As we mentioned above, once a pre-synaptic neuron generates a spike, its synaptic interactions will not take effect on the post-synaptic neurons immediately, but after a specific delay. However, the synaptic delay varies between neurons, leading to poor efficiency for checking whether a synaptic delay falls into the range of current time step. What's worse, in one simulator\cite{thu_snn_1, thu_snn_2}, parallel computing is performed on temporal domain. That is an principled error, because there must be some dependencies between two time steps.\\
Fortunately, an innovative scheduling scheme is implemented to deal with varied synaptic delays. First, synaptic interactions will be reordered according to their delays and the corresponding threads as shown in Fig.~\ref{fig:delay_syn_order}, where deeper in color means higher delay of this synaptic interaction. Then, the synaptic interactions will be performed from lowest synaptic delays to highest synaptic delays according to the current time step. As shown in Fig.~\ref{fig:spk_delay_comp}, the elements with red border are the currently accessing ones in this time step, and the elements in color 'aluminium' are the ones which have been accessed in the past. By this simple but clear arrangement, parallel computation of synaptic interactions at the thread level can be well executed with varied synaptic delays.\\
\subsection{Overlapping of Communication and Computation}
As a typical sparse problem, another bottleneck is the inter-process communication, especially in large-scale simulation. Therefore, a practical way of general improvement is to overlap communication and computation. In this section, we will describe why the overlapping is possible, and our optimized implementation.\\
\subsubsection{Spikes Broadcast and Buffer}
\begin{figure}[h!] 
	\centering
    \includegraphics[width=0.48\textwidth]{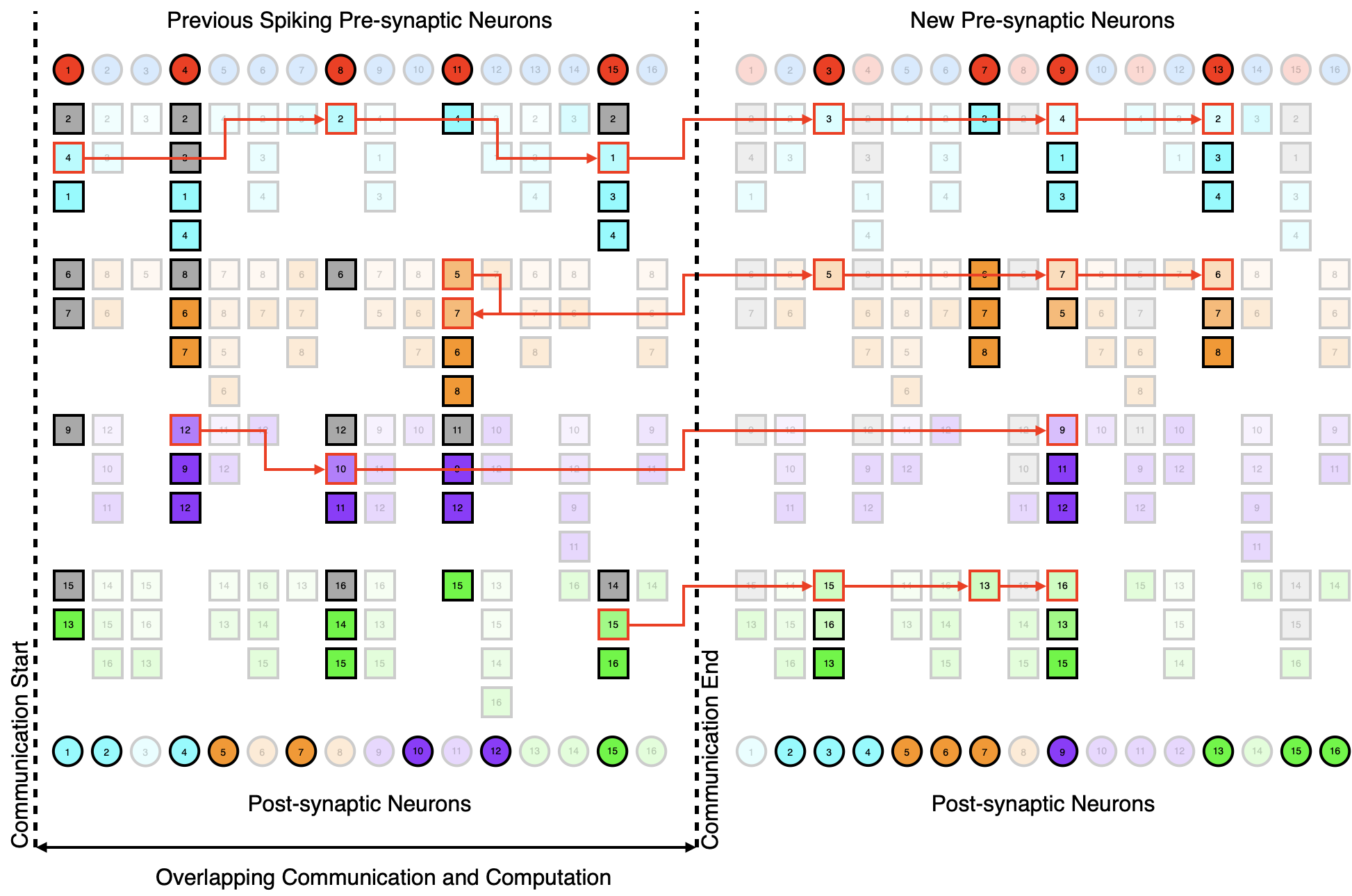}
    \caption{Overlapping of Communication and Computation in One Time Step}
    \label{fig:overlap_comm_comp}
\end{figure}
As we mentioned in (11), (13) and (14), in indegree sub-graphs, the only dependencies are pre-synaptic neurons. And also, the spiking graph $^{in}\!S_{s}$ can be defined by spiking pre-synaptic neurons $^{in}\!V^{pre}_{s}$ only, where the post synaptic neurons $^{in}\!E_{s}$ and synaptic interactions $^{in}\!V_{s}$ can be easily found out in (4). Therefore, the goal of communication is to let all processes know which pre-synaptic neurons generate spikes in each time step, reducing lots of requirement for data transfer between processes. This procedure can be called as Spikes Broadcast, where each process broadcast their spiking pre-synaptic neurons to others.\\
Additionally, as we mentioned before, because of varied synaptic delays, the computation of synaptic interactions from one spiking pre-synaptic neurons might spans several time steps. So, a buffer is required to place these spiking pre-synaptic neurons, until their synaptic interactions are all finished. However, these previous spiking pre-synaptic neurons in buffer make the overlapping become possible. In each process, the synaptic interactions from them can be computed before new spiking pre-synaptic neurons arriving from other processes, to overlap communication and computation, as shown in Fig.~\ref{fig:overlap_comm_comp}.\\
\subsubsection{Dedicated Thread for Communication}
\begin{figure}[h!] 
	\centering
    \includegraphics[width=0.35\textwidth]{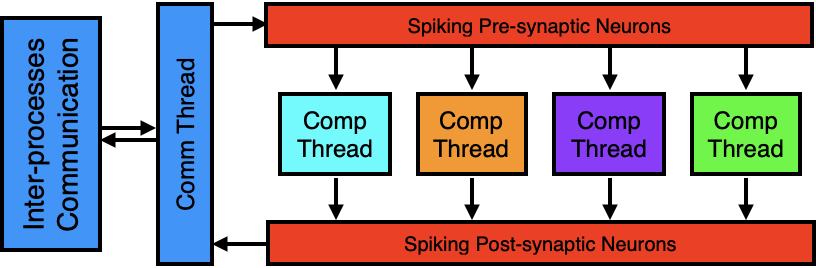}
    \caption{Dedicated Thread for Communication}
    \label{fig:comm_thread}
\end{figure}

In order to overlap communication and computation as much as possible, a dedicated thread for communication has been set up in each process, while others remain unchanged for computation. Different from this implementation \cite{bsc} by using OpenMP Tasks for overlapping, the synaptic interactions of CORTEX are based on the indegree sub-graph of each thread, where the threads for computation should be specific during the entire simulation. As shown in Fig.~\ref{fig:comm_thread}, the dataflow is similar to a circulatory system. Neural dynamics with synaptic interactions will generate spikes from post-synaptic neurons. Then, the communicating thread broadcasts these spikes to other processes. In the destination, these spikes will be received and placed in a buffer as spiking pre-synaptic neurons. After that, all computing threads will fetch these spiking pre-synaptic neurons to perform synaptic interactions and neural dynamics without dependencies at both process and thread level. And thus, there will be new spikes generated from post-synaptic neurons, to start a new loop.\\
\section{VERIFICATION AND EVALUATION}
Because of various modeling methods and biological parameters, it is hard to demonstrate which one is better in performance or scalability without simulating a same brain model. However, it is impossible to perform an apple-to-apple comparison among all simulators, because some of them might not support some kinds of modeling methods, and there should be lots of work to build the same simulation on all existing platforms. So, to perform a fair and meaningful comparison, cases with same modeling method will be simulated by our simulator: CORTEX and NEST Simulator, the SOTA\cite{snn_robot_mouse} with available open-source code. Two cases will be in comparison as follows:
\begin{itemize}
  \item {Verification: A random balanced network with STDP mechanism in nonlinear synaptic dynamics, provided by NEST\cite{hpc_benchmark}.}
  \item {Evaluations: A Multi-scale spiking network model of marmoset in paxinos structural connectome, built from the connectome data at the website: \url{https://connectome.marmosetbrainmapping.org/Paxinos_SC_matrix}, cell density: \url{https://www.marmosetbrain.org/cell_density}, interareal distance: \url{https://analytics.marmosetbrain.org/static/data/marmoset_brain_connectivity_1_0_interareal_distance_matrix.txt} and the internal architecture is derived from\cite{potjans_2014}: \url{https://github.com/nest/nest-simulator/tree/master/pynest/examples/Potjans_2014}}
\end{itemize}
\subsection{Verification}
Each aspect we mentioned above has its own meaning. The first case is a balanced random network in which the excitatory-excitatory neurons exhibit STDP with multiplicative depression and power-law potentiation. The number of incoming synaptic interactions per neuron is fixed and independent of network size. By this case, we want to show that CORTEX is able to support nonlinear synaptic interactions, including complex computation with varied data structures like queue, stack and so on, and still free from data racing without any mutex or atomic operation. Also, the results of thread mapping will be check in this case. Simply, if a edge or post-vertex is accessed by different threads, Abort will be called by CORTEX.\\
As described in NEST\cite{hpc_benchmark}, the neurons' firing rates of this network should be lower than 10 Hz, where the results might differ due to the chaotic nature of the dynamics. On different simulators or runtime platforms, minor differences in the computation of neural and synaptic interactions can lead to completely different spike sequences after a short time\cite{precise_spike}.\\
Although this case is named as "HPC Benchmark" by NEST, we don't use it as a real performance benchmark, because this case just represents a simple theoretical SNN without enough biological properties from brain architecture in reality, especially the varied density and delays of synaptic interactions, upon which many of the optimization efforts we mentioned above are focused.\\
\subsection{Evaluation}
\begin{figure}[h!] 
	\centering
    \begin{subfigure}{0.4\textwidth}
        \centering
        \includegraphics[width=1.1\textwidth]{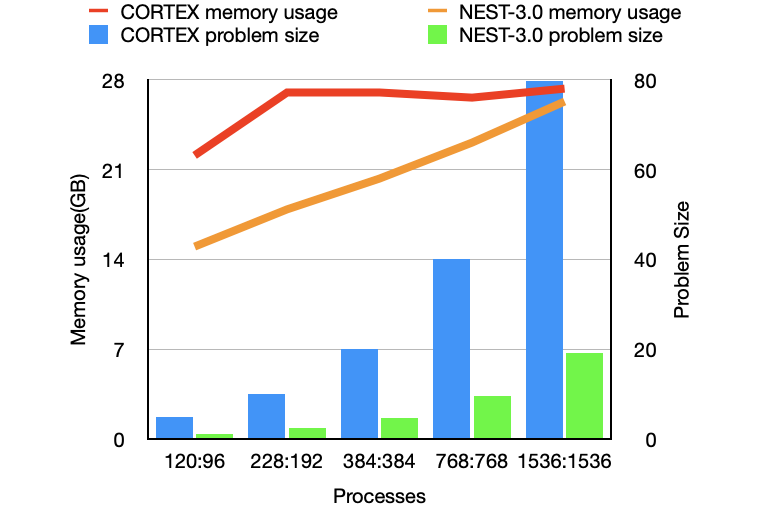}
        \caption{Memory Usage}
        \label{fig:mem_usage}
    \end{subfigure}
    \begin{subfigure}{0.4\textwidth}
        \centering
        \includegraphics[width=1.1\textwidth]{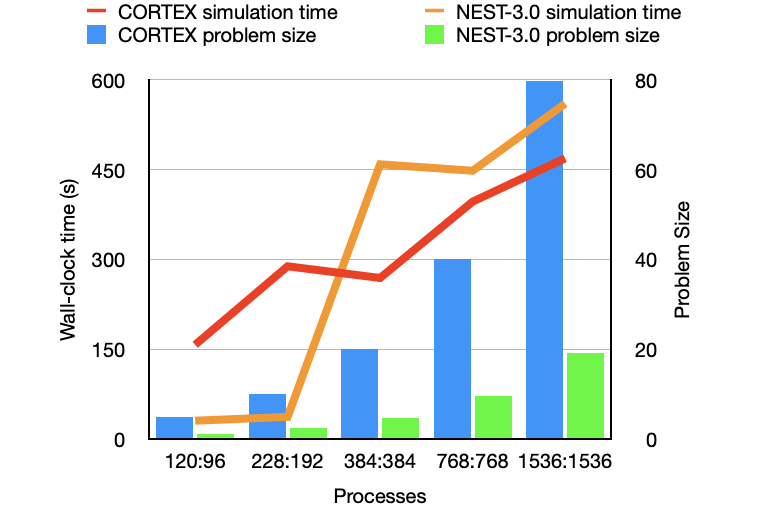}
        \caption{Simulation Time}
        \label{fig:wall-clock_time}
    \end{subfigure}
    \caption{Performance Result: Memory usage is the maximal value of consumption in each node. The normalized problem size 1 contains 1 million neurons with 3.7 billion synaptic interactions. All variables for numerical computing are in IEEE 754 64-bit floating point format without any compression on accuracy.}
    \label{fig:perf_result}
\end{figure}

Using LIF neuron model\cite{iaf_psc_exp_1,iaf_psc_exp_2}, the marmoset cerebral cortex simulation is a typical large-scale SNNs built from biological data, which can show the superiority of CORTEX in terms of not only application-available problem size but also computing performance. In our belief, this superiority of CORTEX is not limited to a specific case as we mentioned above, almost every brain architecture driven by biological data can enjoy CORTEX. Unfortunately, up to now, many existing cases\cite{multi_area_1, multi_area_2, multi_area_3} from other simulators can't run on the Fugaku supercomputer due to too many technical problems with system environment and software dependencies. Also, another reason is that the dynamic response and computing intensity of this brain architecture remain unchanged with different scaling problem sizes. Therefore, it is suitable to be as a baseline, to demonstrate the performance between different problem sizes.\\
Even so, the source code of CORTEX will be available on Github, and everyone is welcome to build any brain architecture on CORTEX and to make comparisons with other simulators on leading-edge supercomputers, where the neuron and synapse templates along with connecting definitions will be updated to support a wide range of modeling methods in future.\\
The performance is measured using the average time spent on the benchmark test as we mentioned above. The comparison between CORTEX and NEST Simulator includes memory consumption and simulation time, exhibiting the general performance in multi-scaling problem size, as shown in Fig.~\ref{fig:perf_result}. The normalized problem size of 1 contains 1 million neurons with 3.8 billion synaptic interactions, and there are 4 processes per node. Also, all variables for numerical computing are in IEEE 754 64-bit floating point format without any compression on accuracy.\\
\subsection{Results and Visualization}
\begin{figure}[h!] 
	\centering
    \begin{subfigure}{0.2\textwidth}
        \centering
        \includegraphics[width=1.2\textwidth]{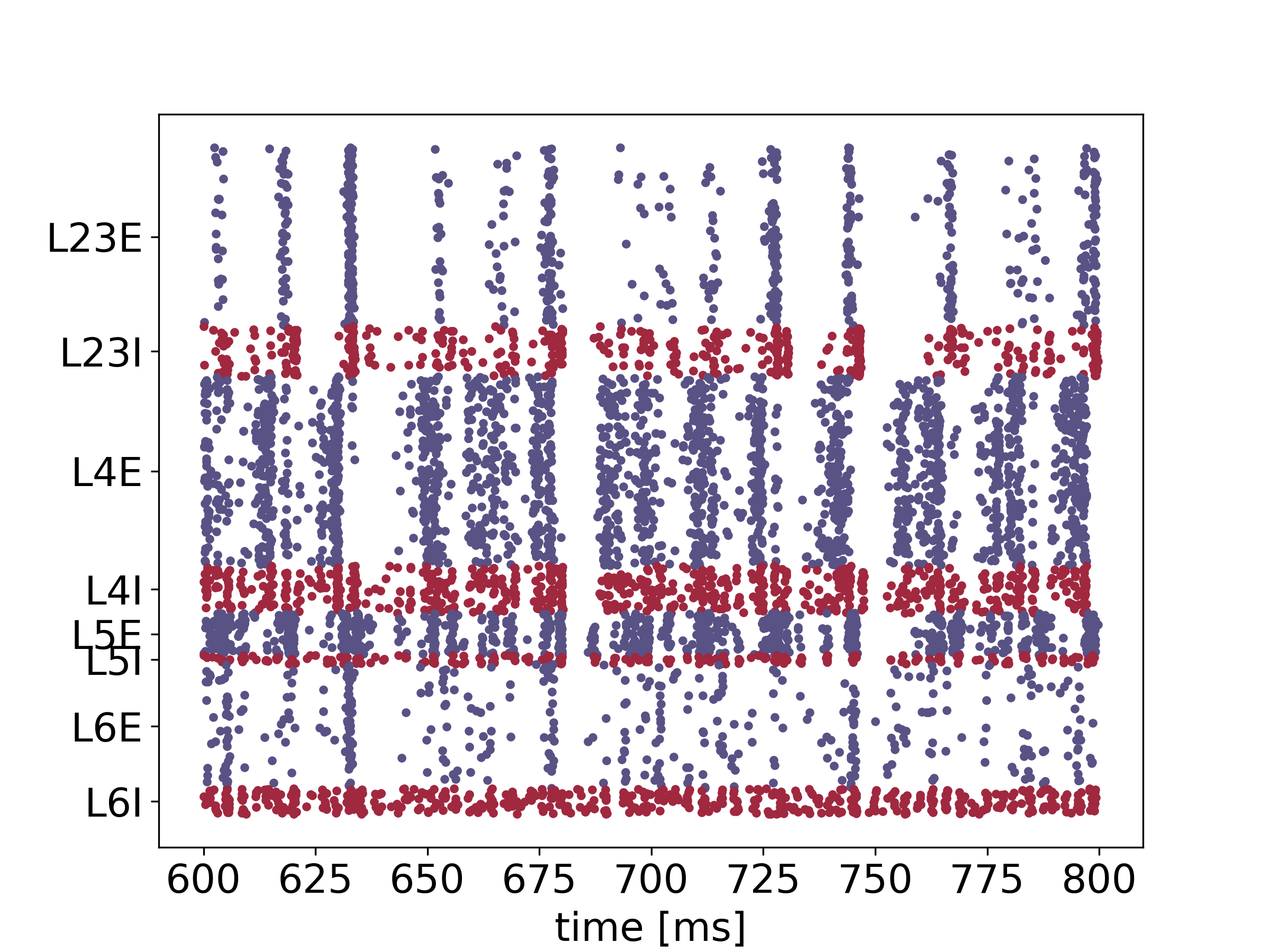}
        \caption{CORTEX Simulator}
        \label{fig:c_raster}
    \end{subfigure}
    \begin{subfigure}{0.2\textwidth}
        \centering
        \includegraphics[width=1.2\textwidth]{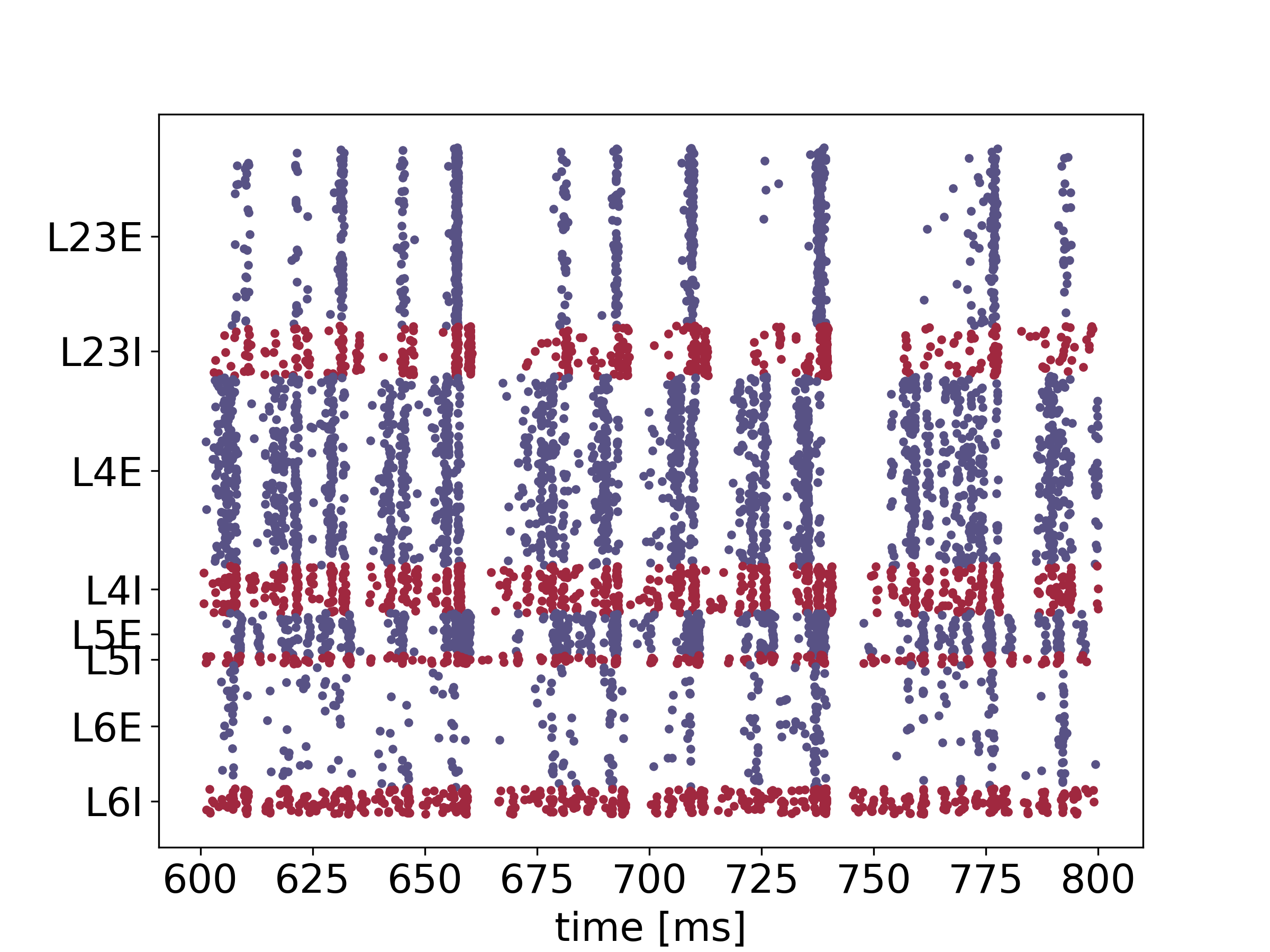}
        \caption{NEST Simulator}
        \label{fig:n_raster}
    \end{subfigure}
    \caption{Rasters Plots of Cortical Activities}
    \label{fig:raster_result}
\end{figure}

Because of the chosen LIF model \cite{iaf_psc_exp_1,iaf_psc_exp_2}, whose computing intensity is much lower than that of other models, the scalability might not be as good as shown in other publications\cite{fdu_brain} using Hodgkin-Huxley model\cite{hh_neuron} with much higher computing intensity. Also, although it is possible to perform larger simulation on CORTEX, the gap of memory consumption and computing performance between CORTEX and NEST Simulator within 1536 processes (384 nodes) has been huge enough.\\
Two raster results of area V1 are shown in Fig.~\ref{fig:raster_result}, which are similar to each other with slight differences, because the computing procedures and random generations between these two simulators are not exactly the same.\\
\section{DISCUSSION}
Despite introducing slightly greater complexities to our algorithmic framework, a key point we wish to highlight is the impact of these innovations on scaling simulation performance and capacity. By maintaining numerical integrity, we overcome the limitations imposed by sparsity. This approach significantly advances our ability to support simulations of the entire human brain, marking a substantial improvement over previous methodologies. Our successful application of CORTEX on the Fugaku Supercomputer demonstrates its potential to handle increasingly larger problems, moving us closer to the ambitious goal of a full human-scale brain simulation, the pinnacle challenge in our field.\\
\subsubsection{Domain Specific Optimizations}
On the practical side, there are still many ways to improve CORTEX at various aspects, especially with domain specific optimizations. In future, CORTEX will be further implemented with optimizations on specific architectures, which are not limited in homogeneous CPU-based systems. Admittedly, implementation on heterogeneous system like CUDA GPU\cite{cuda_gpu} or MN-Core\cite{mn_core} accelerators should be more challenging, directly facing the sparsity of brain architecture in both spatial and temporal domains. Therefore, we expect that further improvements on both the algorithmic and engineering considerations can trigger a breakthrough in future.\\
\subsubsection{Further Improvement on Communication}
In brain simulation, there are lots of irregular spike broadcasts in each time step. Obviously, \texttt{MPI\_Bcast} are not suitable for irregular large amounts of communication. Brain Simulation Broadcast (BSB) is a broadcast acceleration library specifically designed for this communication pattern, which can automatically packs/unpacks spikes into/from messages and adaptively routes the messages among processes to decrease the number of small messages in the physical network. To decrease the communication latency, the interfaces of BSB are designed in a producer-consumer model, and its implementation utilizes Unified Notifiable RMA (UNR) library to achieve synchronization-free communication. We will optimize the communication of CORTEX with BSB in the next update.\\
\subsubsection{Towards Whole Human-Scale Brain Simulation}
As we mentioned at the outset, human-scale brain simulation is one of the most ambitious scientific challenges of the 21st century, where the whole brain includes the cerebellum and basal ganglia, with synaptic plasticity, whose properties are much more different from the architecture of cerebrum cortex only. Within this brain architecture, minor differences in parameters and mechanisms can lead to completely different firing rates, synaptic density and mechanisms, related to differences in computing intensity and memory consumption. Obviously, with such a heterogeneous brain architecture, load balance of both computation and storage is more challenging to achieve at the same time.\\
\section{ACKNOWLEDGMENTS}
This work was supported by Program for Promoting Researches on the Supercomputer Fugaku (hp200139). We are very grateful to Dr. Jun Igarashi and Prof. Kenji Doya.\\

\vspace{12pt}
\end{document}